\documentclass[aps,pre,twocolumn,showpacs,preprintnumbers,amsmath,amssymb,floatfix,showkeys]{revtex4-1}
\usepackage{graphicx}
\usepackage[usenames,dvipsnames]{xcolor}
\usepackage{lineno}
\usepackage{amsthm}
\usepackage{mathtools}
\usepackage{subfigure}
\usepackage{silence}
\usepackage[normalem]{ulem}
\definecolor{winered}{rgb}{0.8,0,0}
\definecolor{darkb}{rgb}{0,0,0.8}
\usepackage[pdftex]{hyperref}
\hypersetup{
	colorlinks=true,
    linkcolor=winered,
    citecolor=blue,
    anchorcolor=black,
    menucolor=red,
    filecolor=magenta,
    urlcolor=darkb
}

\usepackage[draft]{todonotes}   % notes showed

\def\lm{$\lambda_{max}\ $}
\newlength{\dummysp}
\newcommand{\tr}{\mathop{{\hbox{Tr} \, }}\nolimits}

\newcommand{\beq}{\begin{eqnarray}}
\newcommand{\eeq}{\end{eqnarray}}

\newcommand{\gappeq}{\mathrel{\rlap {\raise.5ex\hbox{$>$}}
{\lower.5ex\hbox{$\sim$}}}}
\newcommand{\lappeq}{\mathrel{\rlap{\raise.5ex\hbox{$<$}}
{\lower.5ex\hbox{$\sim$}}}}

\newcommand{\ben}{\begin{enumerate}}
\newcommand{\een}{\end{enumerate}}
\newcommand{\bit}{\begin{itemize}}
\newcommand{\eit}{\end{itemize}}

\newcommand{\bondavg}{{\langle \mathbf{v}_b\rangle}}
\newcommand{\siteavg}{{\langle \mathbf{v}_s\rangle}}

\def\[{\left [}
\def\]{\right ]}
\def\({\left (}
\def\){\right )}
\def\lc{loop$\ $}
\DeclareMathOperator{\Tr}{Tr}

\begin{document}

\title{Examples of renormalization group transformations for image sets}

\author{Samuel Foreman}
\email{samuel-foreman@uiowa.edu}
\affiliation{Department of Physics and Astronomy, The University of Iowa, Iowa City IA 52242}
\affiliation{Computational Sciences Division, Argonne National Laboratory, Argonne, IL 60439 USA} 
\author{Joel Giedt}
\email{giedtj@rpi.edu}
\affiliation{Department of Physics, Applied Physics and Astronomy,
Rensselaer Polytechnic Institute, 110 8th Street, Troy NY 12180 USA}
\author{Yannick Meurice}
\email{yannick-meurice@uiowa.edu}
\affiliation{Department of Physics and Astronomy, 514 Van Allen Hall, The University of Iowa, Iowa City IA 52242}
\author{Judah Unmuth-Yockey}
\email{jfunmuth@syr.edu}
\affiliation{Department of Physics, Syracuse University, Syracuse, NY 13244 USA}

\date{\today}

\begin{abstract}
Using the example of configurations generated with the worm algorithm for the
two-dimensional Ising model, we propose  renormalization group (RG)
transformations, inspired by the tensor RG, that can be applied to sets of images. We relate criticality to
the logarithmic divergence of the largest principal component.  We discuss the changes
in link occupation  under the RG transformation, suggest ways to obtain data
collapse, and compare with the two state tensor RG approximation near the fixed
point. 
\end{abstract}
\maketitle

\section{Introduction}
%\todo[inline]{These are todonotes, in order to prevent them from showing up,
%comment out line 31 and uncomment line 30 in the tex file.}
%\comment{These are comments, in order to prevent them from showing up, comment
%out line 35 and uncomment line 34 in the tex file}

Machine learning is a general framework for recognizing patterns in data
without detailed human elaboration of the rules for doing so.  As an example,
a very general function, with many parameters (for example, thousands or
millions) can be optimized on a training set, where the desired output is known.
The problem is typically nonconvex and plagued by over-fitting problems, and so advanced methods are necessary in order to get reliable answers.   One tool that has been
exploited is principal component analysis (PCA), which reduces the
dimensionality of the data to the most important ``directions.''  
Immediately
the practitioner of renormalization group (RG) methods recognizes an
analogy, since the RG techniques are also supposed to identify the
most important directions in an enlarged space of Hamiltonians.
One of the motivations of the present research is to make this analogy
more concrete.

A number of papers \cite{LiWang,Beny,MehtaSchwab} 
attempt to draw a connection between deep learning
and the renormalization group as it appears in physics.  
However, the analogies between renormalization
group flow and depth in a neural network would be strengthened
if one could determine conditions under which fixed points can be identified. 
 It would be helpful to
show more explicitly how passing from one level to another in a neural network genuinely translates to a renormalization group transformation.
There have been steps in the direction of making a full connection.
For instance in \cite{Beny}, the principle of {\it causal influence}
is emphasized.  That is, when descending in depth, only neighboring
nodes should influence the outcome of a lower level node.  We have
also implemented this in a simple training scheme in earlier
work \cite{foreman2017}.  It can be called ``cheap learning'' because
far fewer variational parameters are involved, due to the constraints
of locality.
% This can be viewed as a type of real space renormalization group. (Judah)
In \cite{MehtaSchwab} it is emphasized that deep neural networks
outperform shallower networks for reasons which may ultimately
be understood in terms of the power of the renormalization group.
Other topics related to machine learning, such as principal
component analysis \cite{BraddeBialek} have been previously
interpreted in terms of the renormalization group (in this
case momentum shells). Machine learning has also been used to identify phase transitions in numerical simulations \cite{2017NatPh..13..431C,PhysRevB.94.195105,2017PhRvE..95f2122H,2017PhRvE..96b2140W}.

RG transformations are
usually defined in a space of couplings/Hamiltonians, but typically, it is not
possible to write down  Hamiltonians directly associated with images sets.
In this article we propose RG transformations that can be applied to a specific
set of images but which could be generalized for other image sets, and can also be understood analytically without any graphical
representation. We use the well-studied example of the two-dimensional Ising
model on a square lattice.  The spin configurations generated with importance
sampling provide images with black and white pixels. They have features that
can be used to attempt to recognize the temperature used to generate them. % there should be a citation here right?
However, constructing blocked Hamiltonians in configuration space is a
difficult task which involves approximations that are difficult to improve. In
other words, it is very difficult to explicitly construct the exact RG transformation
mapping the original couplings among the Ising spins into coarse grained ones. 

A better control on the RG transformation can be achieved by using the tensor
renormalization group (TRG) method
\cite{PhysRevLett.99.120601,PhysRevB.79.085118,PhysRevB.86.045139,prb87,prd88,prd89,pre89}.
The starting point for this reformulation is the character expansion of the
Boltzmann weights which is also used in the duality transformation
\cite{RevModPhys.52.453}. This leads to an exact expression of the partition
function  as a sum over closed paths which can be generated with importance
sampling using  the worm algorithm \cite{prok87} and then pixelated. These
samples will be our sets of images indexed by the temperature used to generate
them.  The procedure is reviewed in Sec.~\ref{configs_as_images}. 

The goal of a RG analysis is to study systems with large correlation lengths in
lattice spacing units and iteratively replace them by coarser ones with a
larger effective lattice spacing. This process is useful if we can tune a
parameter such as the temperature towards its critical value.  Typical image
sets such as the MNIST data can be thought as ``far from criticality'' and the
use of RG methods for such a data set may be of limited interest
\cite{foreman2017}.  Criticality may sometimes refer to the choice of
parameters used in data analysis \cite{PhysRevD.83.105014}. % do we mean to use refer here?
It seems crucial
to introduce a systematic way to deal with the concept of criticality in ML. 

The PCA is a standard method to analyze sets of images.  In configuration
space, the PCA analysis is identical to the study of the spin-spin correlation
matrix. In particular, the largest eigenvalue \lm  is directly connected to the
magnetic susceptibility which diverges at criticality
\cite{PhysRevB.94.195105}. In the loop representation (worms), we will show
that \lm diverges logarithmically at criticality  with a constant of
proportionality which can be estimated quite precisely ($3/\pi$).  This is
explained in Sec.~\ref{sec:pca}.  More generally, it seems reasonable to
identify the criticality with the divergence of \lm.

The advantage of rewriting the high-temperature expansion in terms of tensors
is that it allows a very simple blocking (coarse-graining) procedure where a
group of sites is replaced by a single site.  In the TRG approach the blocking
procedure is local. This leads to simple and exact coarse-graining formulas
because we can separate the links into two categories: one half of the links
are inside the blocks and integrated over while the other half are kept fixed
and ensure the communication among the blocks \cite{prb87}. The main goal of
this article is to relate blocking procedures that can be applied to sets of
pixelated images, to approximate TRG transformations.  A short summary of the
TRG procedure is given in Sec.~\ref{sec:trg}.

Having defined criticality, the next step is to define a RG transformation for
sets of ``legal'' loop configurations, also called ``worm configurations" later, sampled at various temperatures.  In
Sec.~\ref{sec:rgimages},  we propose a family of transformations which replaces
two parallel links in a block by a single link carrying a specific value $x$. We call this procedure $1+1\rightarrow x$ hereafter. 
In the case $1+1\rightarrow 0$, the blocked
images follow the same rules (for legal configurations) as the original ones.
There is a clear analogy with the 2-state approximation for the TRG method which is
described in the rest of the paper. In the 2-state TRG approximation, the average fraction
of occupied links show a characteristic crossing at a critical point and a
collapse when the distance to the critical point is appropriately rescaled at
each iteration. The average fraction of occupied links in the blocked worm configurations (with
$1+1\rightarrow 0$) show a somewhat similar behavior in the low temperature
phase. However, on the high-temperature side, we observe a ``merging'' rather
than a crossing. In Sec.~\ref{sec:collapse},  we provide explanations for the
similarities and differences between the two procedures.

In
Sec.~\ref{sec:nbtrg} we discuss an approximate 2-state TRG method to calculate
the average number of bonds through several iterations. 
\begin{figure}[htpb]
    \centering 
    \includegraphics[width=8.6cm]{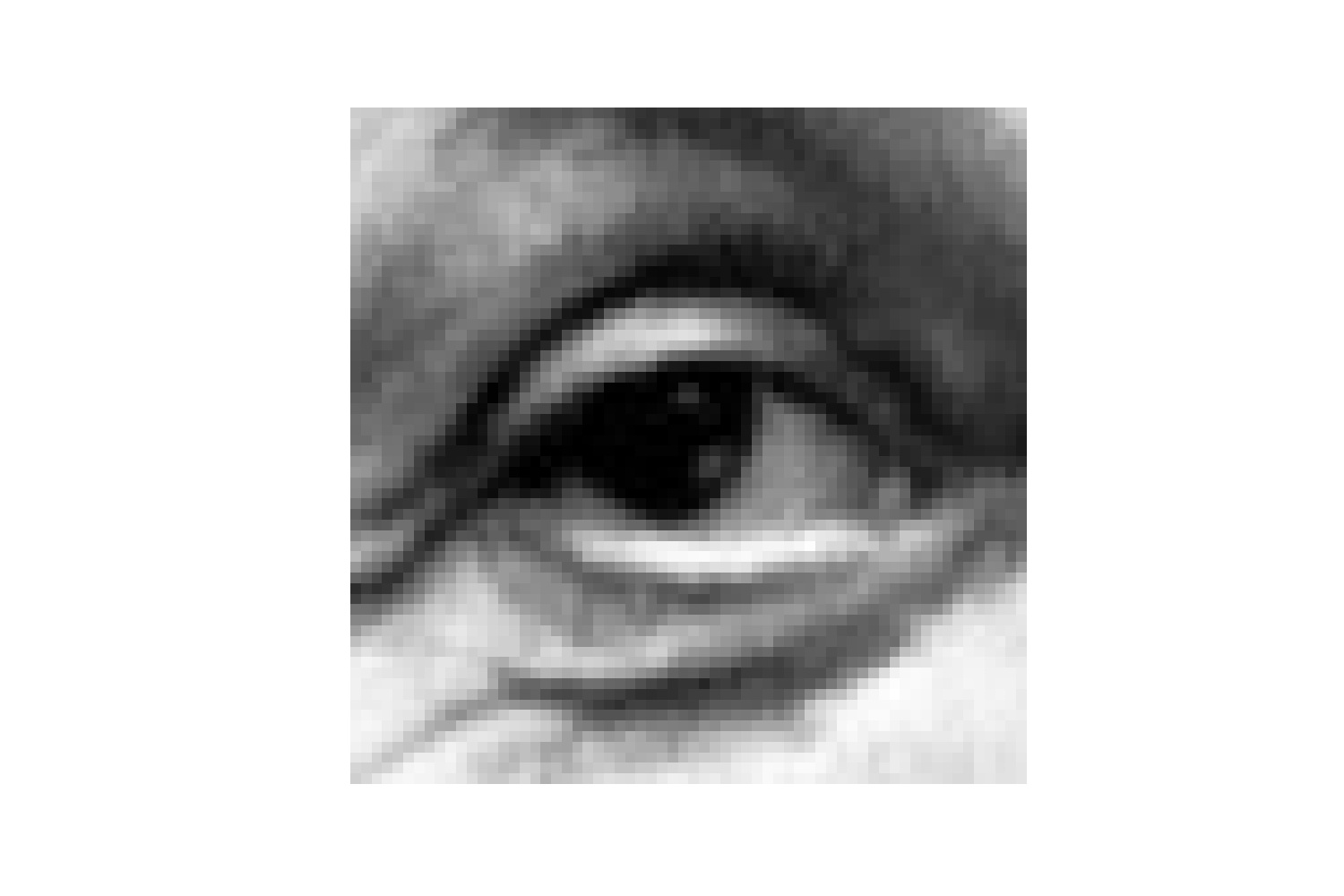}
    \includegraphics[width=8.6cm]{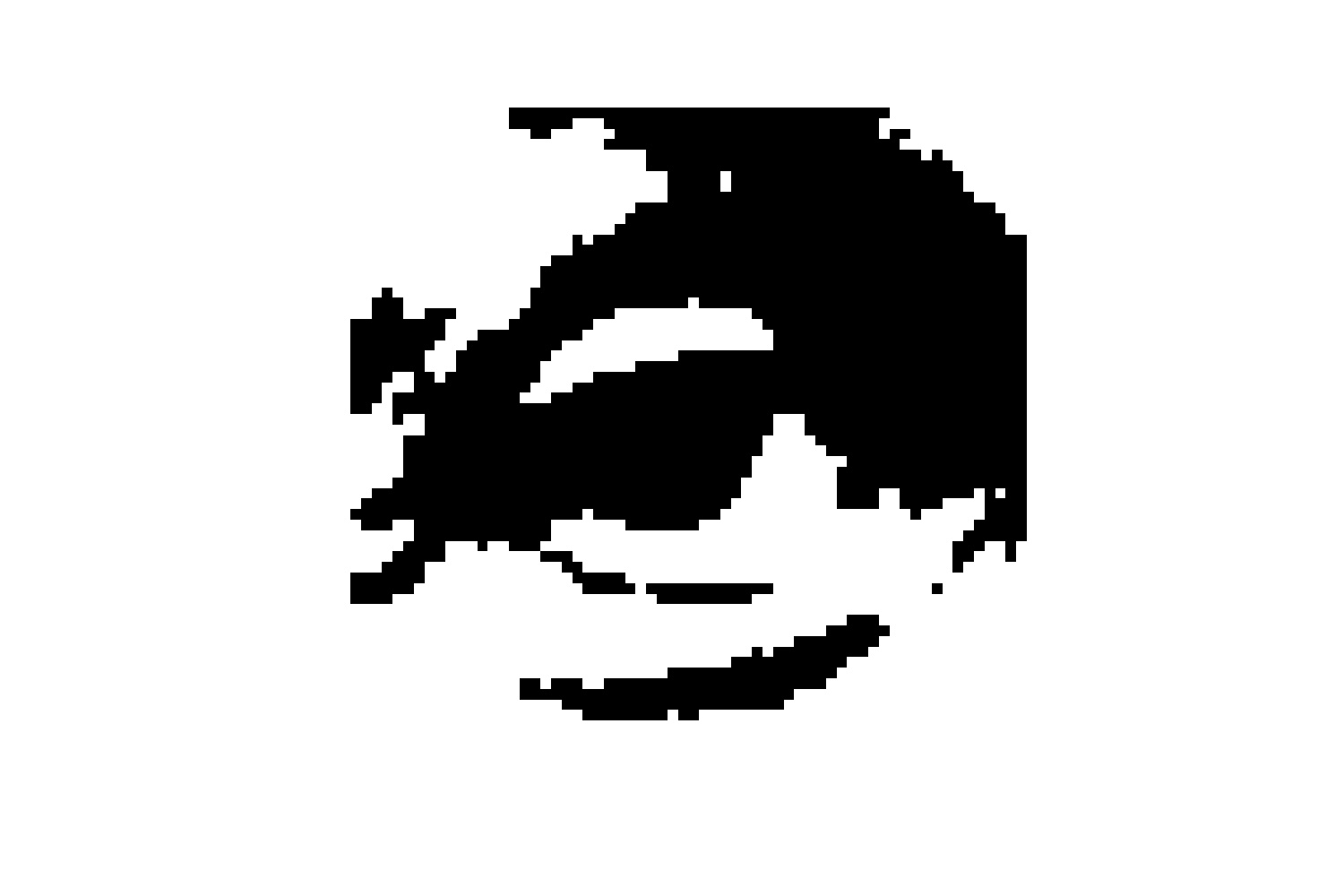}
    \includegraphics[width=8.6cm]{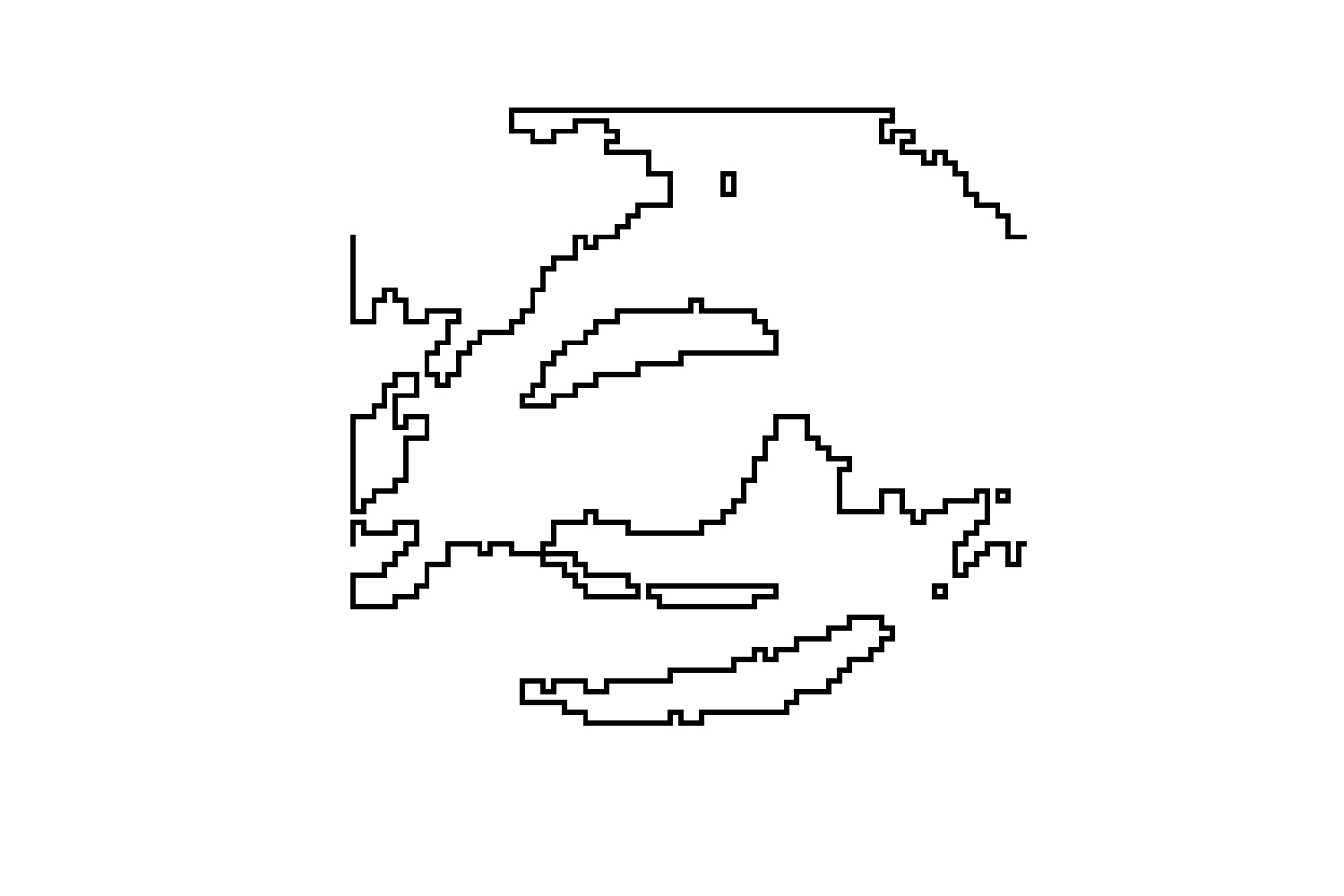}
    \caption{\label{fig:eye}Picture of an eye with 4096 pixels; black and white
    version with a graycut at  0.72; boundaries of the black domains.}
    \label{default}
\end{figure}
The worm configurations can be directly connected to spin
configurations using duality  \cite{RevModPhys.52.453}: they are the boundary
of the positive spin islands. This suggests that the methods discussed here
could be applied to generic images. Boundaries of generic grayscale pictures
can be defined by converting the picture to black and white pixels.  A grayscale picture with gray values between 0 and 1 can be converted into an Ising
spin configuration, by introducing a ``graycut'' below which the value is
converted to 0 (spin down) and above which the value is converted to 1 (spin up). It is then possible to construct the boundaries of the spin up domains.
This is illustrated in Fig.~\ref{fig:eye}. Possible applications are discussed
in Appendix~\ref{sec:cifar}. 

\section{From loops to images}
\label{configs_as_images}

In the following we consider the two-dimensional Ising model with spins
$\sigma_i = \pm 1$ on a square lattice.  The partition function reads
\begin{equation}
Z = \sum_{ \{ \sigma_i \} } e^{\beta \sum_{ \langle i, j \rangle} \sigma_i \sigma_j}
\label{ih}, 
\end{equation}
where $\langle i, j \rangle$ denotes nearest neighbor sites on the square lattice. 
In some occasions we will use the notation $T=1/\beta$ for the temperature. 
The partition function can be rewritten by using the character
expansion~\cite{prokofevIsing04}
\begin{equation}
\exp(\beta\sigma)=\cosh(\beta)+\sigma\sinh(\beta),
\end{equation}
and integrating over the spins. Factoring out the $\cosh(\beta)$, each link can
carry a weight 1 when unoccupied or $t\equiv \tanh(\beta)$ when occupied. The
integration over the spins guarantees that an even number of occupied links is
coming out of each site \cite{RevModPhys.52.453}. The set of occupied links
then form a ``legal graph" with $N_b$ occupied links. The partition function
can then be written as sum over such legal graphs. If $\mathcal{N}(N_b)$
denotes the number of legal graphs with $N_b$ links we can write: 
\begin{equation}
Z=2^V(\cosh(\beta))^{2V}\sum_{N_b} t^{N_b}\mathcal{N}(N_b)
\label{eq:nbsum}
\end{equation}
Using the fact that $\tanh(\beta)=\exp(-2\tilde{\beta})$, with $\tilde{\beta}$
the inverse dual temperature, Eq.~(\ref{eq:nbsum}) has the same form as a
spectral decomposition using a density of states and a Boltzmann weight (with
$2N_b$ playing the role of the energy).  Details of this reformulation can be found in Appendix~\ref{ssec:loop_representation}.

As shown in Appendix \ref{ssec:heat_capacity}, we can use derivatives of the logarithm of the
partition function to relate $\langle N_b\rangle$ to the average energy, and the
bond number fluctuations,
\begin{equation}
\label{eq:fluc}
    \langle \Delta_{N_b}^2 \rangle \equiv \langle\left(N_b - \langle
    N_b\rangle\right)^2\rangle,
\end{equation}
to the specific heat per site. From the logarithmic singularity of the specific
heat we find that 
\begin{equation}
    \langle \Delta_{N_b}^2\rangle /V =-\frac{2}{\pi}\ln(|T-T_c|)+{\rm regular}.
    \label{eq:specific_heat_fluctuation_eq2}
\end{equation}
In the following we use interchangeably the ``bond'' terminology, for instance in
$N_b$ as in \cite{prokofevIsing04}, and the  link terminology more common in
the lattice gauge theory context.  In all our numerical simulations we use
periodic boundary conditions which guarantees translation invariance.
% \comment{Periodic boundary conditions are used for all numerical simulations
% in order to guarantee translational invariance. (rewrite?)}

We will show in Sec.~\ref{sec:trg} that the  new form of the partition function
in Eq.~(\ref{eq:nbsum}) can also be written in an equivalent way as a sum of
products of tensors with four indices contracted along the links of the lattice.

The contributions to Eq.~(\ref{eq:nbsum}) can be sampled using a worm algorithm
\cite{prok87} outlined in Appendix \ref{ssec:monte_carlo_implementation}.  Using this algorithm, we generated
multiple configurations at each temperature ($N_{configs} \approx10,000$) which
are then used for averaging. For example, we can calculate the average number
of occupied bonds at a particular temperature by averaging over all
configurations.

\begin{figure}[htpb]
    \centering 
    \includegraphics[width=5.73cm]{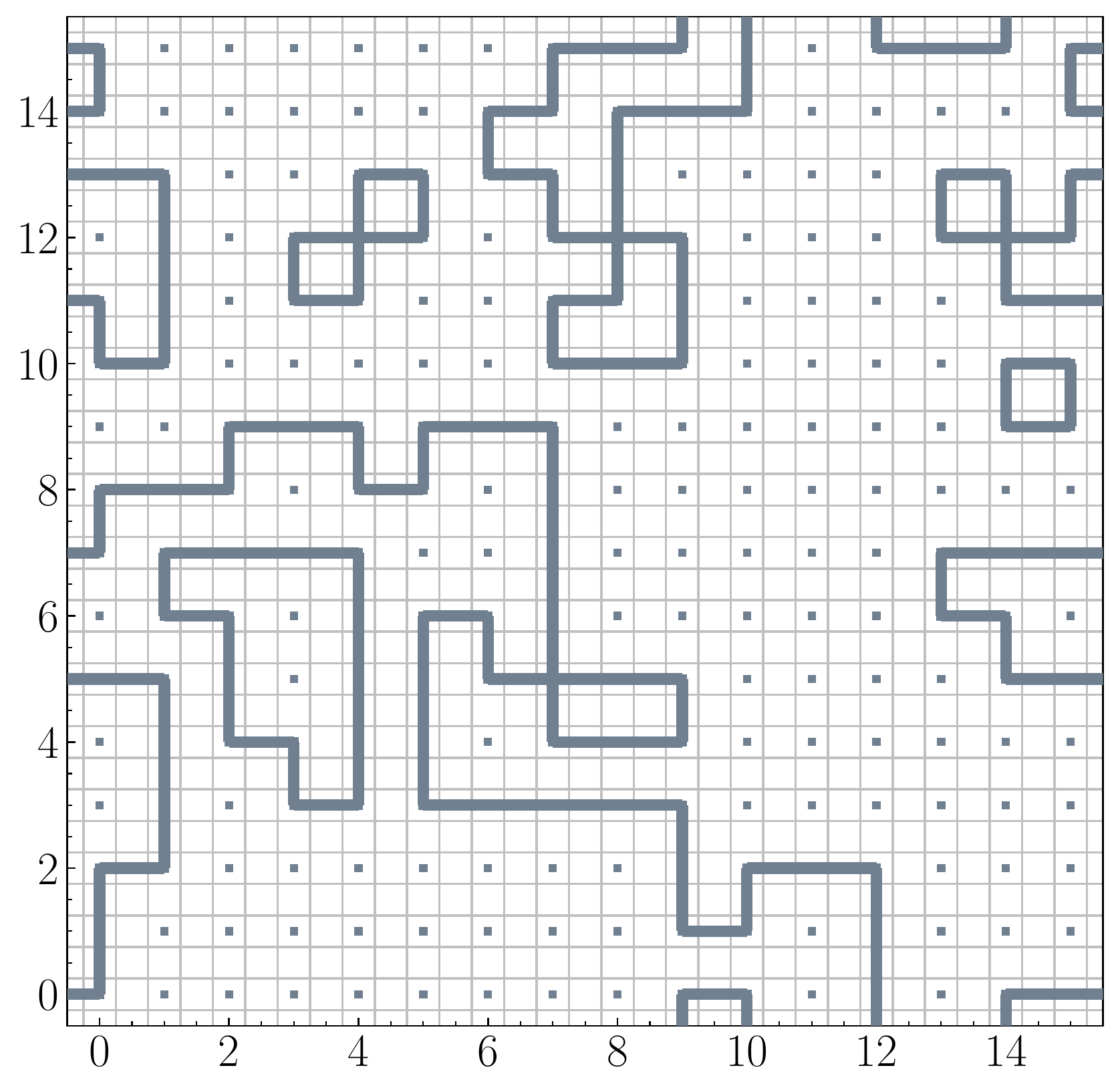}
    \includegraphics[width=5.73cm]{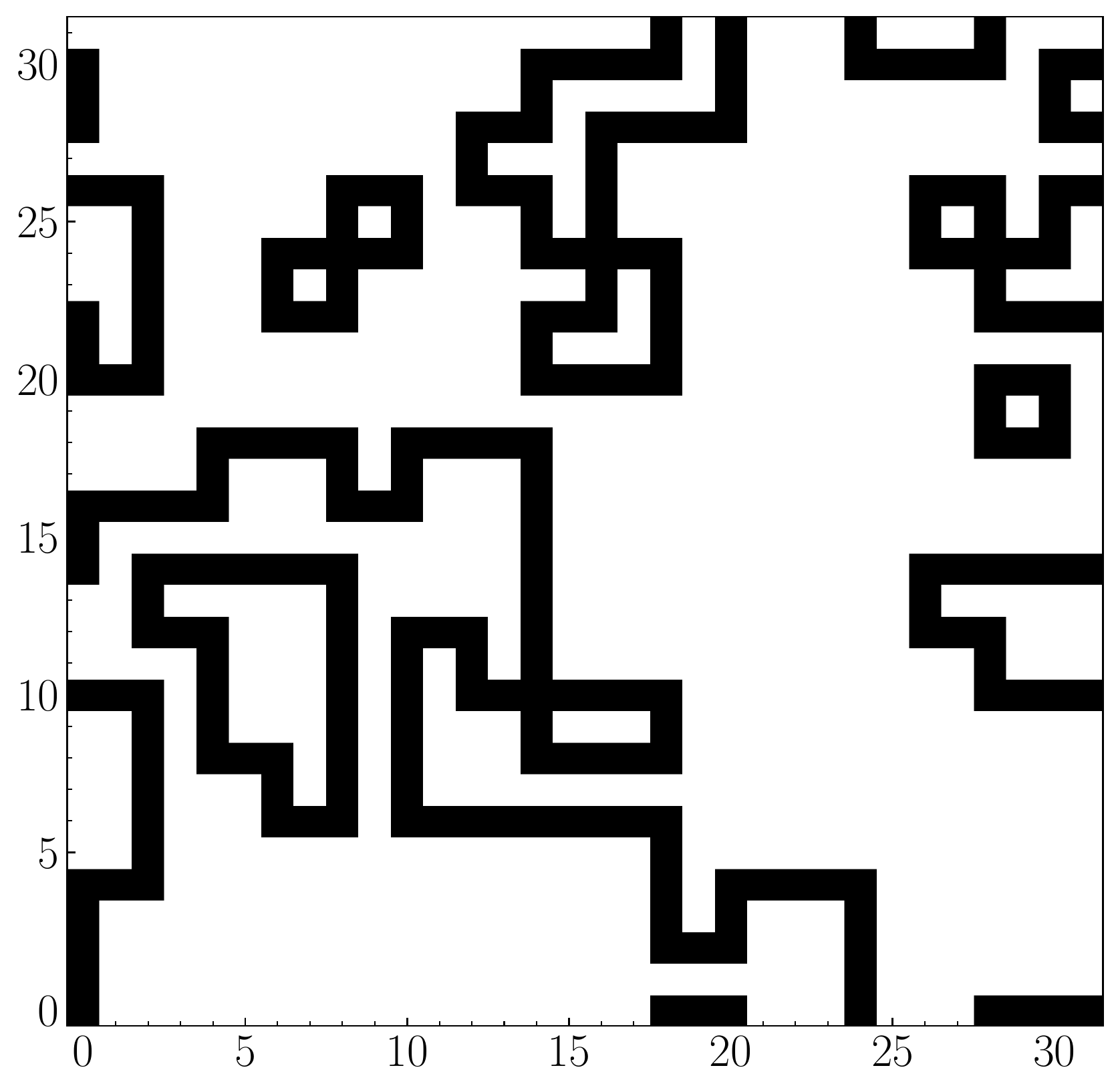}
    \caption{Legal worm configuration on an $L \times L$ lattice with periodic
        boundary conditions (top) and its equivalent representation as a
    $2L\times2L$ black or white pixels (bottom).}
    \label{fig:worm_as_image}
\end{figure}

Using a legal graph (worm configuration), we can construct an image by
introducing a lattice of $2L\times2L$  pixels with a size of one half lattice
spacing.  One quarter of these pixels are attached to the sites, one quarter to
the horizontal links and one quarter to the vertical links. The remaining
quarter are in the middle of the plaquettes and always white.  In this
representation, each site, link, and plaquette are designated an individual
pixel, where occupied links and their respective endpoints are colored black.
An example of this representation is shown in Fig.~\ref{fig:worm_as_image}.
We can then flatten each of these images into a vector $\mathbf{v} \in
\mathbb{R}^{4V}$, with $v_i \in \{0, 1\}$. This allows us to write the number
of occupied bonds, in a single configuration, $N_b$ as
\begin{equation}
    \sum_{j = bonds} v_j = N_b
    \label{eq:link_sum}
\end{equation}

\section{PCA and criticality}
\label{sec:pca}

Having now sets of images for a range of temperatures, we can apply PCA \cite{Bishop}.
PCA isolates the ``most relevant'' directions in the dataset.  PCA is simply the computation of the eigenvalues
$\lambda_\alpha$ and eigenvectors $u_\alpha$ of the covariance matrix for a
dataset with $N$ configurations corresponding to a given temperature $\{ {\bf
v}^n \}_{n=1}^N$:
\begin{equation}
S_{ij} = \frac{1}{N} \sum_{n=1}^N (v_{i}^n - {\bar v}_i)
(v_{j}^n - {\bar v}_j).
\end{equation}
In this equation, each sample ${\bf v}_j$ is a vector in $\mathbb{R}^{4V}$,
labeled by the indices $i,j = 1,\ldots,4V$.  The PCA extracts solutions to
\begin{equation}
S u_\alpha = \lambda_\alpha u_\alpha
\end{equation}
and orders them, in descending magnitude of $\lambda_\alpha$, which are all
non-negative. The usefulness of PCA is
that one can approximate the data (see for instance the discussion in
\cite{Bishop}) by the first $M$ principal components. 

Illustrations of the PCA for the MNIST data can be found in Sec. 4 of Ref.~\cite{foreman2017}, where we show the eigenvectors corresponding to the
largest eigenvalues and the approximation of the data by subspaces of the
largest eigenvalues of dimensions 10, 20 etc. 
%\todo[inline]{Maybe rewrite the following sentence?}

\begin{figure}[htpb]
    \centering
    \includegraphics[width=8.6cm]{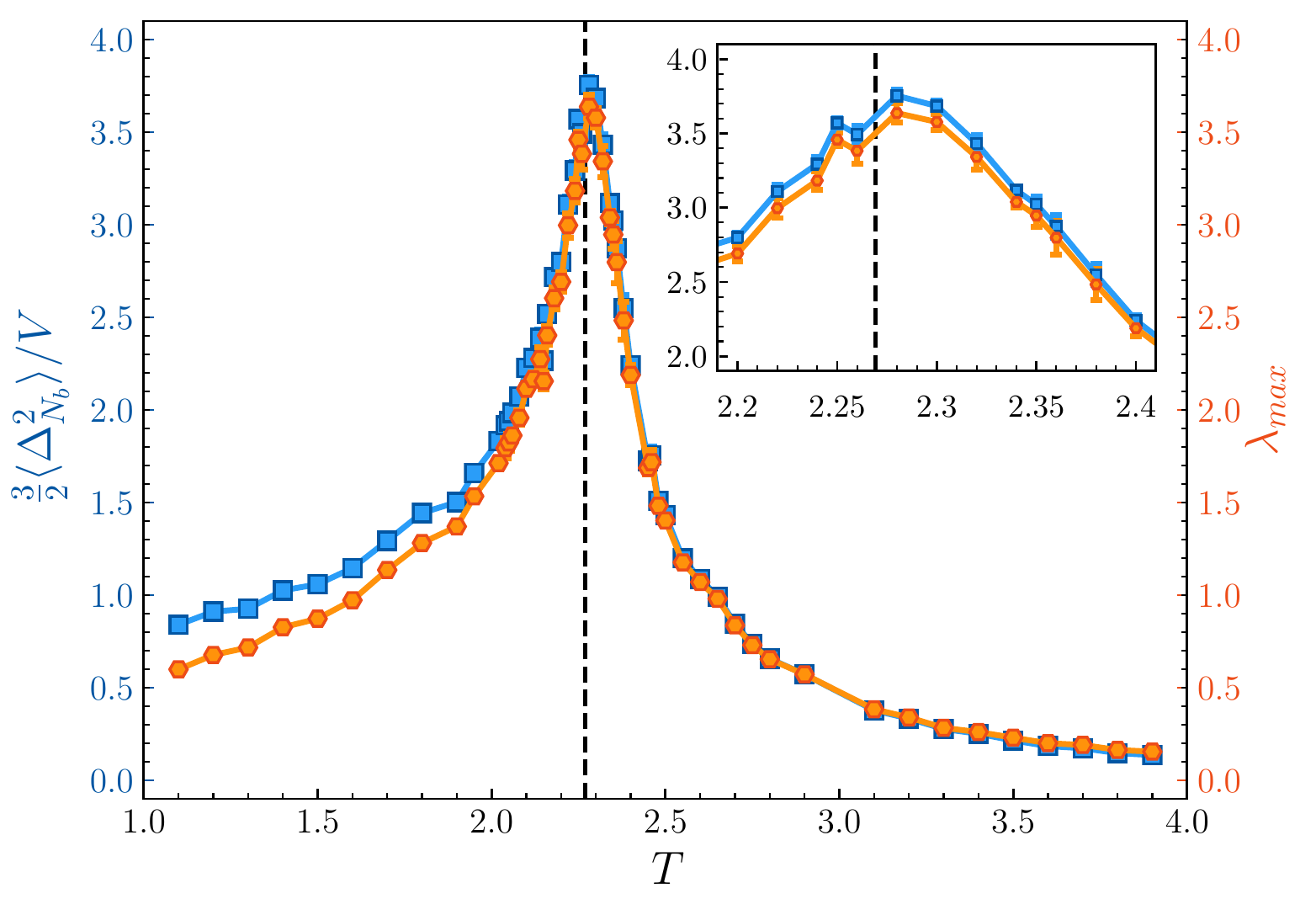}
    \caption{$\lambda_{max}$ and $\frac{3}{2}\left \langle \Delta_{N_b}^2
        \right\rangle / V$ vs. $T$, illustrating the relation between the
        eigenvalue corresponding to the first principal component and the
        logarithmic divergence of the specific heat. The inset shows a
    qualitative agreement near the critical temperature.}
    \label{fig:eigval_fluctuations_unblocked}
\end{figure}

It should be noticed that the PCA is an analysis that can be performed for each
temperature separately and not obviously connected to the closeness to criticality. However, we were able to find a relation between the
largest PCA eigenvalue denoted $\lambda_{max}$ and the logarithmic divergence
of the specific heat, namely
\begin{equation}
    \lambda_{max} \simeq \frac{3}{2}\left \langle \Delta_{N_b}^2
    \right\rangle/V\simeq -\frac{3}{\pi}\ln(|T-T_c|).
    \label{eq:eigval_delta_Nb}
\end{equation}
This property was found by an approximate reasoning shown in Appendix \ref{ssec:heat_capacity} and
relies on two assumptions.  The first one is that the eigenvector associated
with $\lambda_{max}$ is proportional to $\langle \mathbf{v}\rangle$ which is
invariant under translation by two pixels in either direction. The second
assumption is that in good approximation we can neglect the contributions from
sites that are visited twice (four occupied links coming out of one site).
Numerically, only 4\% of sites are visited twice near the critical
temperature which justifies the second assumption.
Fig.~\ref{fig:eigval_fluctuations_unblocked} provides an independent
confirmation of the approximate validity of Eq.~(\ref{eq:eigval_delta_Nb}). 

\section{TRG coarse-graining}
\label{sec:trg}

So far we  have sampled the legal graphs of the high temperature expansion of
the Ising model using the worm algorithm.  An alternative approach is to use a
tractable real-space renormalization group method known as the TRG
\cite{PhysRevB.86.045139,prb87,prd88,prd89,pre89}. 

In order to understand what we want to accomplish by blocking the \lc
configuration, it is useful to first understand the evolution of a tensor
element using the TRG method. 

The tensor formulation used here connects easily with  the worm formulation
used in this paper.  After the character expansion has been carried out, one is
left with new integer variables on the links of the lattice with constraints on
the sites which guarantee the sum of the link variables associated with that
site is even.  Therefore we build a tensor using this constraint and the
surrounding link weights.  The tensor has the form
\begin{align}
    T^{(i)}_{x x' y y'}(\beta) = 
    &\left[\tanh(\beta)\right]^{(n_{x}+n_{x'}+n_{y}+n_{y'})/2}\\ 
    &\times \delta_{n_{x}+n_{x'}+n_{y}+n_{y'}, 0\text{ mod }2}.
\end{align}
% \begin{eqnarray}
%     T^{(i)}_{x x' y y'}(\beta)& = &[\tanh(\beta)]^{(n_{x}+n_{x'}+n_{y}+n_{y'})/2}\\
%     &\  &\delta_{n_{x}+n_{x'}+n_{y}+n_{y'}, 0\text{ mod }2}.
% \end{eqnarray}
Here the notation being used is that this tensor is located at the
$i$\textsuperscript{th} site of the lattice, $n_{\hat{\mu}}$ is the integer
variable, taking value 0 or 1, on an adjacent link, and the Kronecker delta,
$\delta_{i,j}$ is understood to be satisfied if the sum is even.  By
contracting these tensors together in the pattern of the lattice one recreates
the closed-loop paths generated by the high-temperature expansion and exactly
matches those paths which are sampled by the worm algorithm.

Using these tensors one can write a partition function for the Ising model that
is exactly equal to the original partition function,
\begin{equation}
Z =2^V (\cosh (\beta))^{2V}\Tr \prod_{i}T^{(i)}_{xx'yy'} 
\end{equation} 
where $\Tr$ means contractions (sums over 0 and 1) over the links. 

\begin{figure}[htpb]
    \centering
	\includegraphics[width=0.25\textwidth]{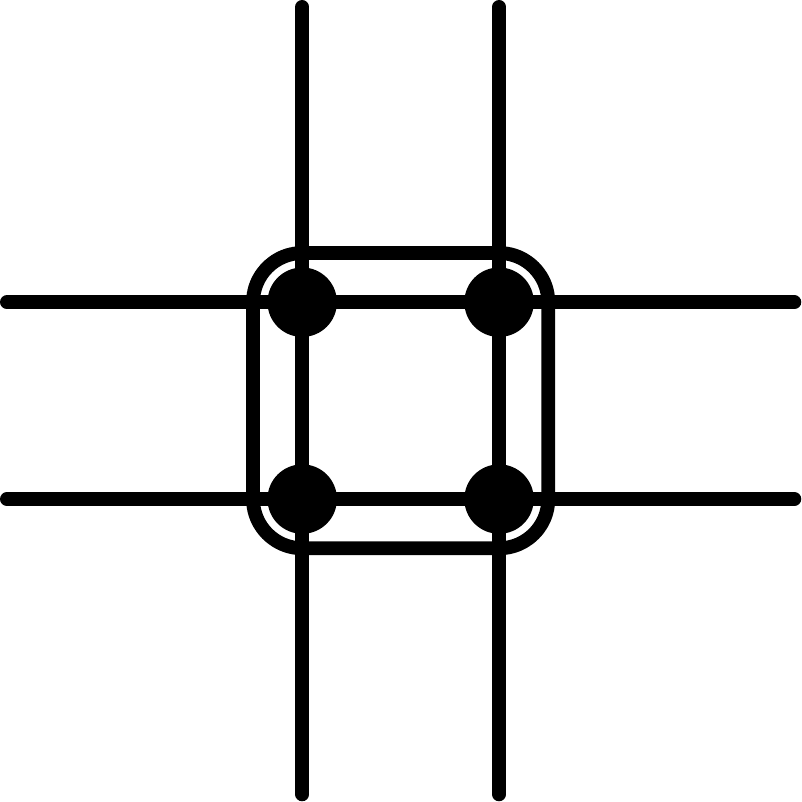}
    \caption{Illustration of the tensor blocking discussed in the text.}
    \label{fig:unit_block}
\end{figure}

\begin{figure}[htpb]
    \centering 
     \includegraphics[width=8.6cm]{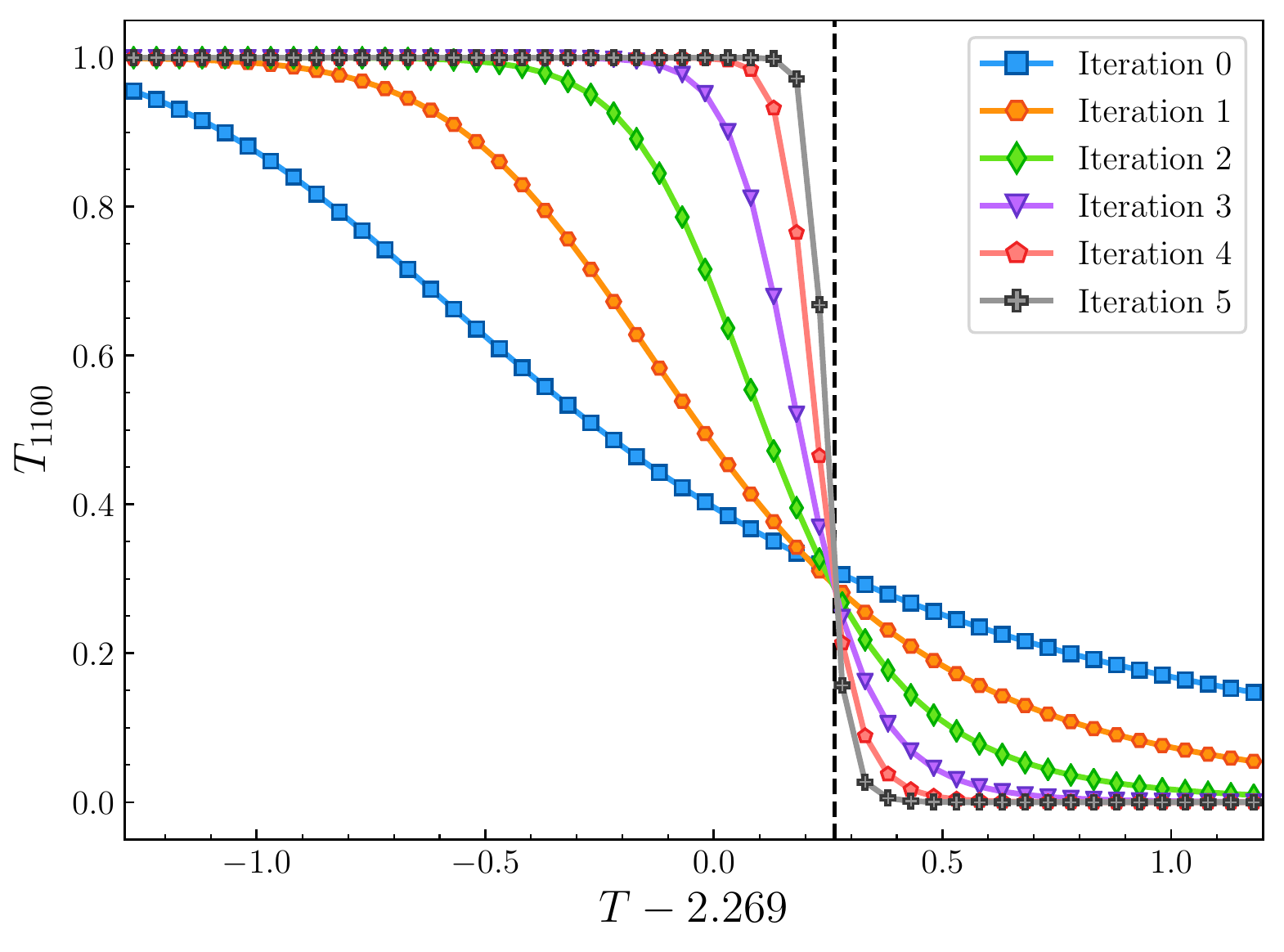}
     \includegraphics[width=8.6cm]{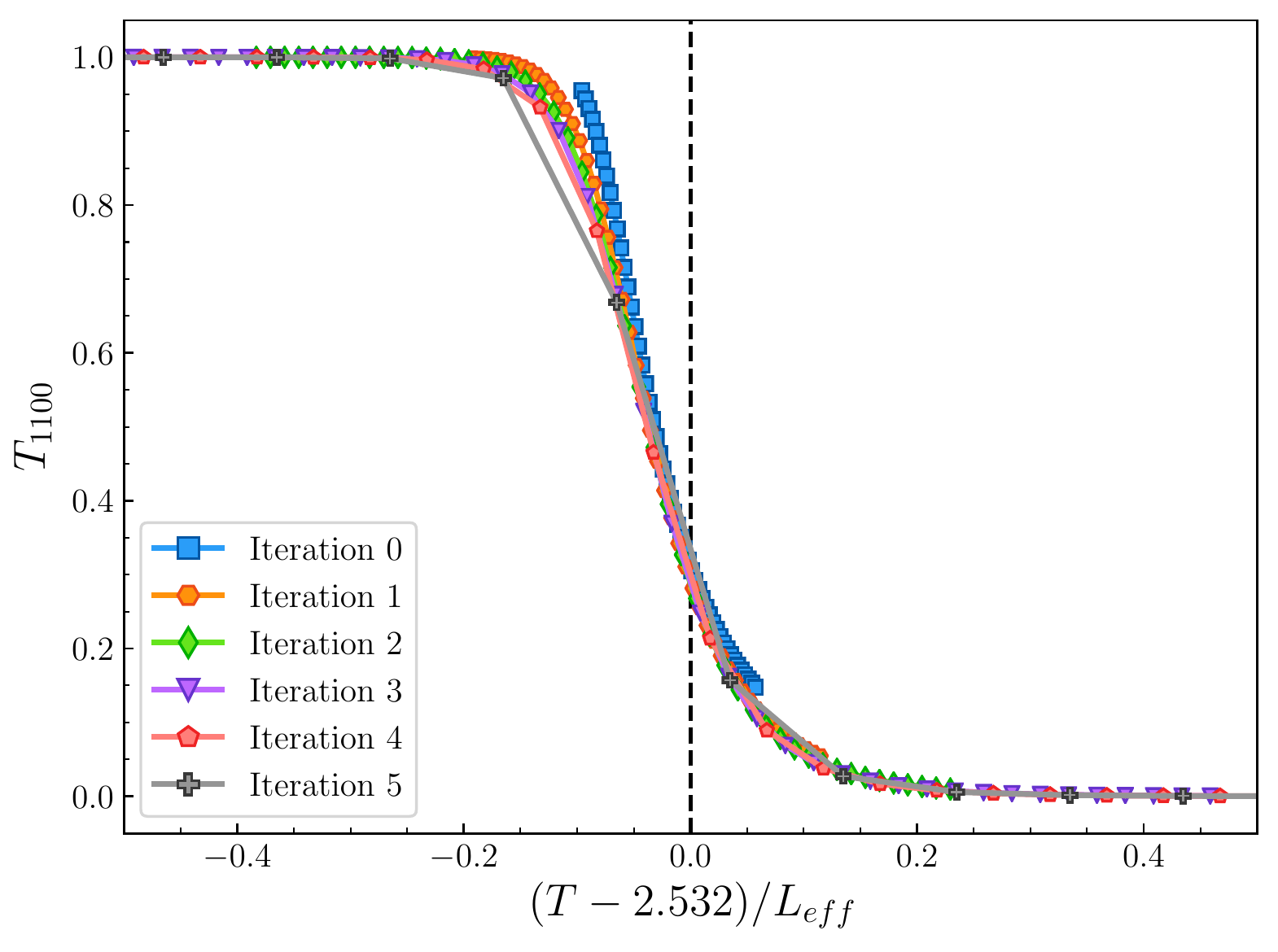}
     \caption{$T_{1100}$ vs $T - T_c$ for six successive iterations of the
         blocking transformation, beginning with an initial lattice $L = 64$
         (top); $T_{1100}$ vs. $(T - T_{c}^{(2s)}) / L_{eff}$ illustrating the
         data collapse, where $T_{c}^{(2s)}$ is the critical temperature of the
         two state projection, beginning at iteration 0 on an $L = 64$ lattice.
     (bottom).}
    \label{fig:T1100}
\end{figure}

The most important aspect of this reformulation is that it can be
coarse-grained efficiently.  The process is illustrated in
Fig.~\ref{fig:unit_block} where four fundamental tensors have been
contracted to form a new ``blocked'' tensor.  This new tensor has a squared
number of degrees of freedom for each new effective index. The partition
function can be written exactly as 
\begin{equation*}
    Z= 2^V(\cosh (\beta))^{2V}\Tr\prod_{2i}T'^{(2i)}_{XX'YY'} \ , 
    \label{eq:ZP}
\end{equation*}
where $2i$ denotes the sites of the coarser lattice with twice the
lattice spacing of the original lattice. 
In practice, this exact procedure cannot be repeated indefinitely and
truncations are necessary. This can be accomplished by projecting the product
states into a smaller number of states that optimizes the closeness to the
exact answer.  A two-state projection is discussed in \cite{prb87} and will be
followed hereafter. Note that in this procedure, $T_{0000}$ is factored out and
the final expression for the other blocked tensors are given in these units.
For definiteness we consider $T_{1100}$ which in the microscopic formulation is
the weight associated with an horizontal line in a \lc  configuration. By
looking at the fixed point equation~\cite{prb87} , one can see that that there
is a high temperature fixed point where all the tensor elements except for
$T_{0000}$ are zero and and a low temperature fixed point where all the tensor
elements are one. In between these two limits, there is a non-trivial fixed
point illustrated by the crossing of iterated values of $T_{1100}$ in
Fig.~\ref{fig:T1100}. Note that because of the two-state approximation, the
critical temperature $T_c$ is slightly higher than the exact one \cite{prb87}.
To be completely specific, the exact $T_c$ for the original model is
$2/\ln(1+\sqrt{2})=2.269 ..$ while for the two state projection with the second
projection procedure of Ref.\cite{prb87}, it is $1/0.3948678 =2.53249...$

It is easy to relate the properties of the iterated curves near the non-trivial
fixed point using the linear RG approximation.  Below we just state the results,
for details and references see \cite{prb87}.  With the blocking procedure used, the scale factor is $b=2$. The eigenvalue in the relevant direction is
$\lambda=b^{1/\nu}=2$ since $\nu=1$.  In Fig.~\ref{fig:T1100}, on can see that
near $T_c$, the height with respect to the crossing point nearly doubles each
time, making the slope twice bigger each time. A nice data collapse can be
reached by offsetting this effect by a rescaling by      $\lambda$=2 the horizontal axis each
iteration as shown in  the bottom part of Fig.~\ref{fig:T1100}. In numerical calculations, we start 
with a finite $L$ (64 in Fig. \ref{fig:T1100}) and then after 
$\ell$ iteration, we are left with an effective size $L_{eff}=L/b^\ell$. 

The remainder of the paper will be dedicated towards obtaining data collapse
for $\langle N_b \rangle$ calculated with successive blockings.

\section{Image coarse-graining}
\label{sec:rgimages}

In an attempt to explicitly connect the ideas from RG theory to similar
concepts in machine learning, we will implement a coarse-graining procedure
directly on the images but in a way inspired by the TRG construction of
Sec.~\ref{sec:trg}. The construction relies on visual intuition and will be
reanalyzed in the TRG context in Sec.~\ref{sec:nbtrg}.

\begin{figure}[htpb]
    \centering
    \includegraphics[width=0.25\textwidth]{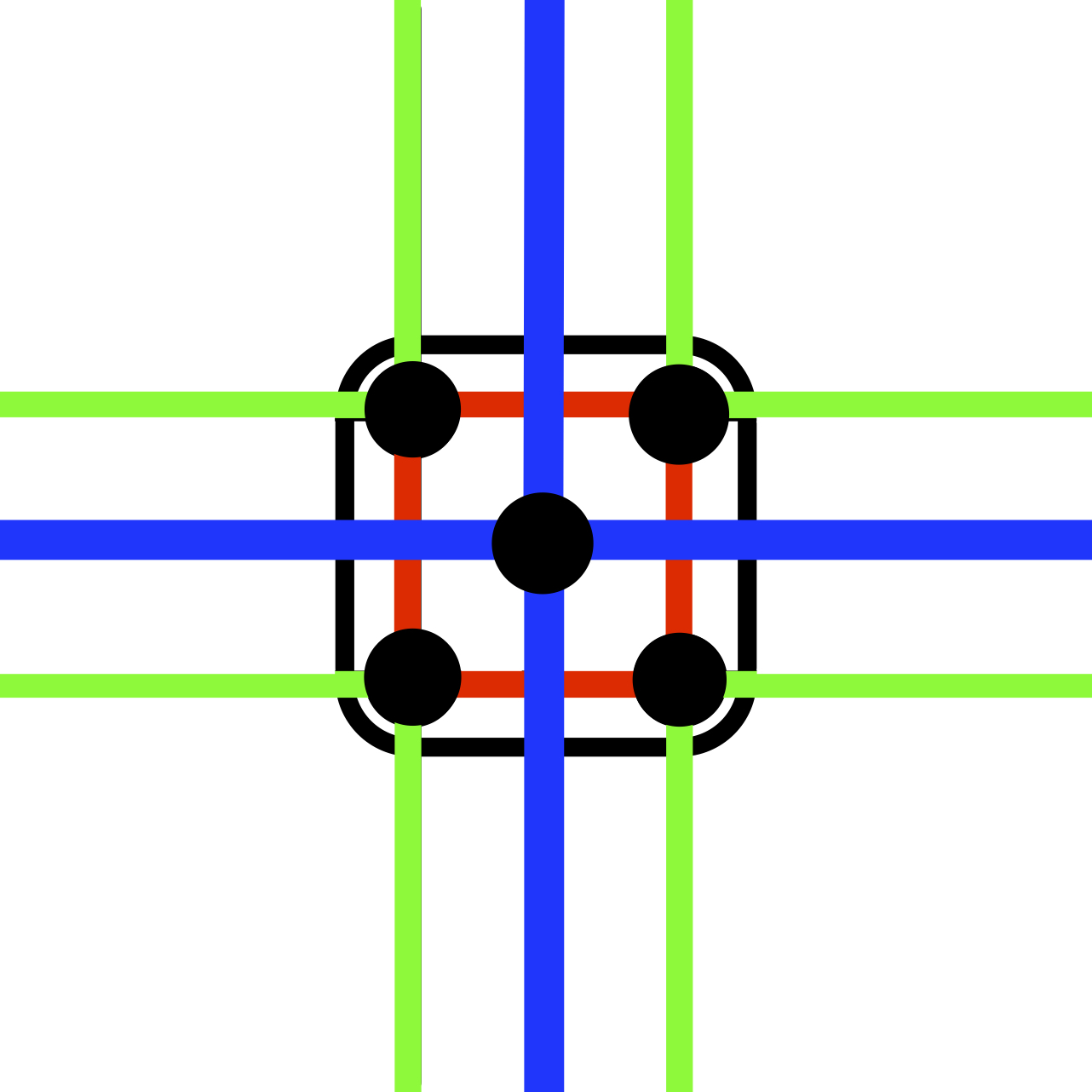}
    \caption{Illustration of an elementary block in the image consisting of
        four sites, four internal bonds (red), eight external bonds (green), and
    four blocked external bonds (blue).}
    \label{fig:blocked-2}
\end{figure}

\begin{figure*}[htpb]
    \centering
    \includegraphics[width=0.31\textwidth]{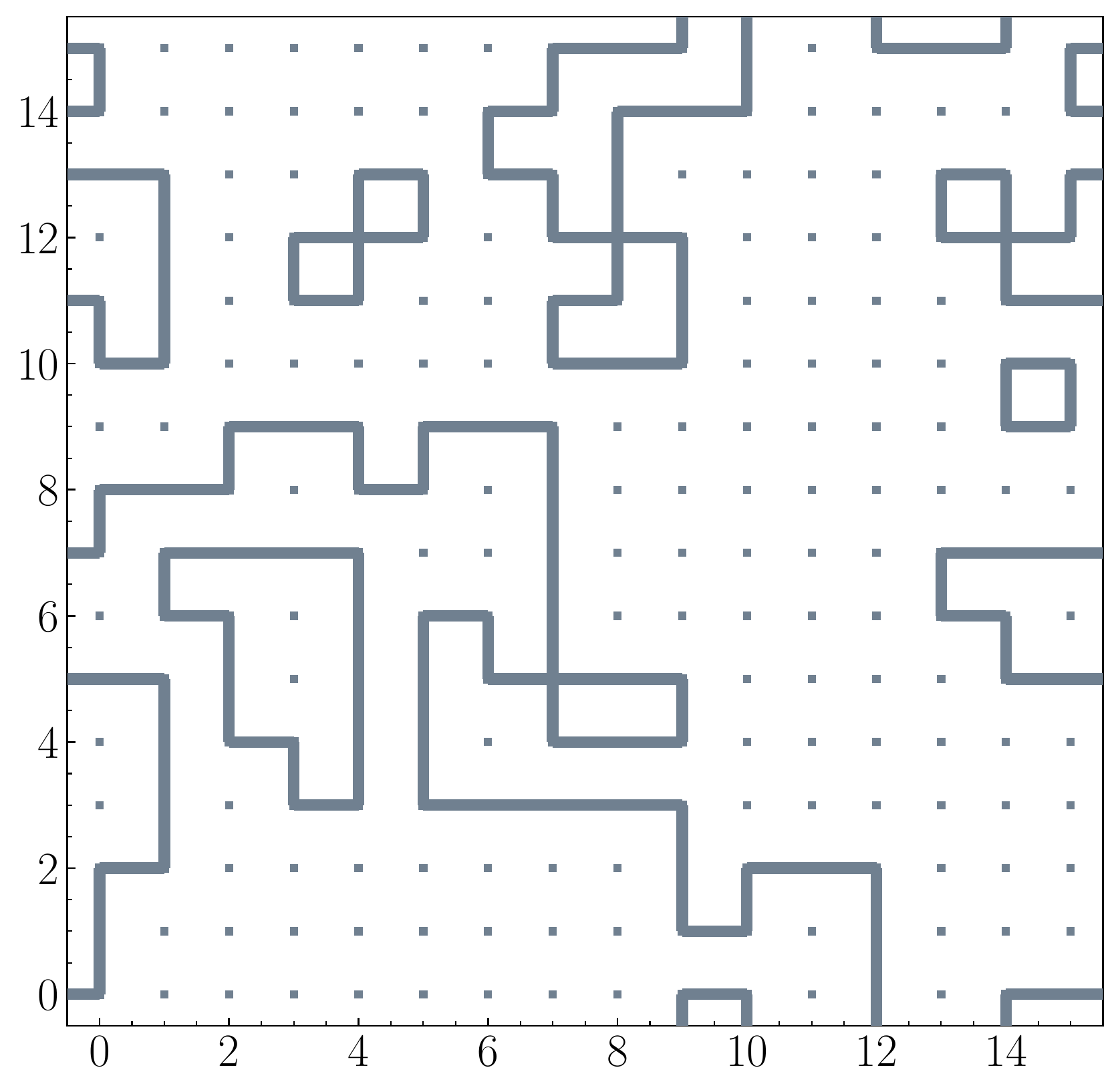}\hfill%
    \includegraphics[width=0.31\textwidth]{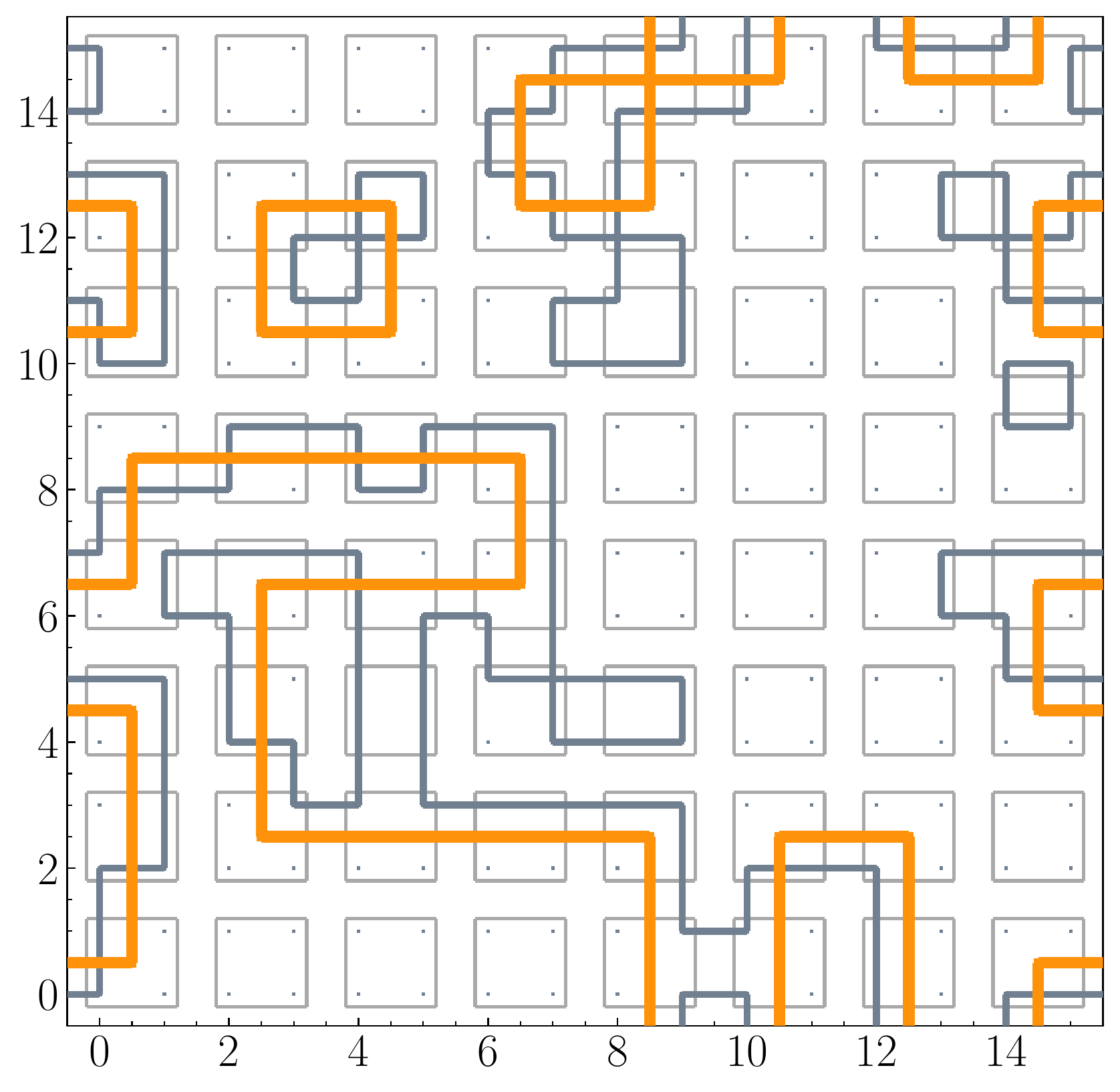}\hfill%
    \includegraphics[width=0.31\textwidth]{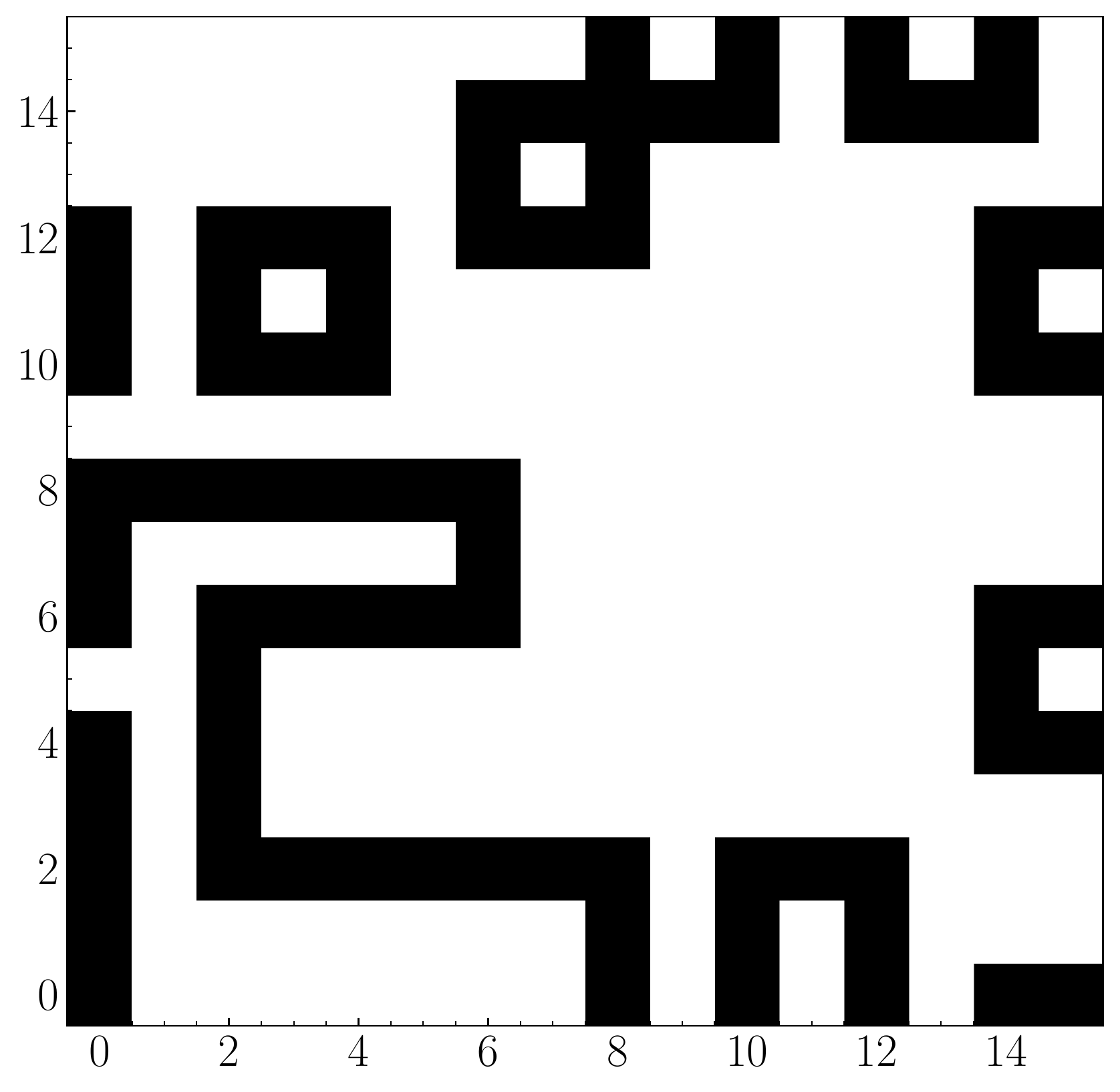}
    \caption{\label{fig:110} Illustration of the $1+1\rightarrow 0$ blocking
        procedure discussed in the text: original configuration (left);
        introduction of the blocks and replacement of single or double bounds
        according to the $1+1\rightarrow 0$ rule (middle); construction of the
    corresponding  blocked image (right).}
\end{figure*}

% As in the TRG coarse graining we divide the original into blocks of $2\times2$
% squares, reducing the size of each linear dimension by a factor of two.
As in the TRG coarse-graining procedure, the image is first divided up into
blocks of $2\times2$ squares, as shown in Fig.~\ref{fig:blocked-2}.
Each of these $2\times2$ squares are then replaced, or ``blocked'', by a single
site with bonds determined by the number of occupied external bonds in the
original square.  In doing so, we reduce the size of each linear dimension by a
factor of two, resulting in a new blocked configuration whose volume is
one-quarter the original. In particular, if a given block has exactly one
external bond in a given direction, the blocked site retains this bond in the
blocked configuration. This seems to be a natural choice. However, if a given
block has exactly two external bonds in a given direction, we can consider
several options. The simplest option is to neglect the double bond entirely, and
we denote this blocking scheme by ``$1+1\rightarrow 0$''. This approach
respects the selection rule (conservation modulo 2) and has the advantage of
maintaining the closed-path restriction. In other words with the
$1+1\rightarrow0$ option, the blocked image corresponds to a legal graph and
the procedure can be iterated without introducing new parameters. This
procedure is illustrated for a specific configuration on a $16\times16$ lattice
in Fig.~\ref{fig:110}.

Alternatively, we can include this double bond in the blocked configuration,
and give it some weight $m$ between 0 and 2.  The examples of $m=$ 1 and 2 are
denoted ``$1+1\rightarrow 1$'', and ``$1+1\rightarrow 2$'' respectively and are
shown in Fig.~\ref{fig:double_bond_weights}. This blocking procedure introduces
new elements and iterations require more involved procedures. This is not discussed 
hereafter. 

\section{Partial data collapse for blocked images}
\label{sec:collapse}
\begin{figure}[htpb]
    \centering 
    \includegraphics[width=8.6cm]{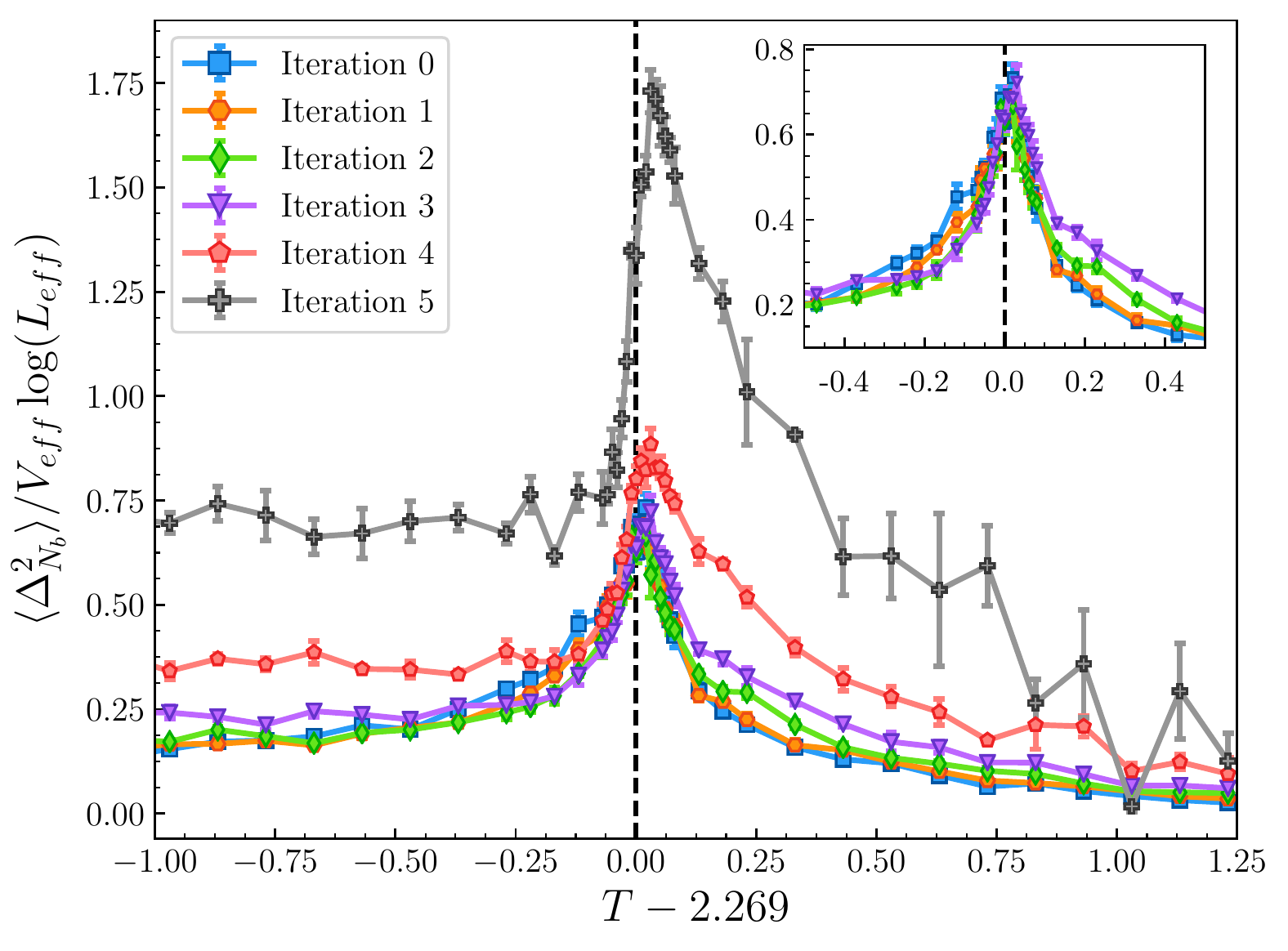}
    \caption{Fluctuations in the average number of bonds $\langle
        \Delta_{N_b}^2\rangle$ vs. temperature $T$ under iterated blocking
        steps beginning with an initial lattice size of $L = 64$. The results
        are scaled by $1 / V_{eff}\log(L_{eff})$ in order to demonstrate the
        data collapse near the critical temperature. This collapse is
        especially apparent in the inset, which shows the results under the
        first three blocking steps, with $L_{eff} = 64, 32, 16$, and $8$.}
    \label{fig:delta_Nb_iterated}
\end{figure}

In this section, we study the properties of $\langle N_b\rangle$  obtained for
successive blockings with the $1+1\rightarrow0$ rule starting with
configurations on a $64\times64$ lattice.  A first observation is that the
$1+1\rightarrow0$ blocking preserves the location of the peak of the
fluctuations $\langle\Delta_{N_b}^2\rangle$. In addition it is possible to
stabilize this quantity for a few iterations by dividing by
$L_{eff}\ln(L_{eff})$. This is illustrated in Fig. \ref{fig:delta_Nb_iterated}.
However, a very different scaling appears for the last two iterations which may
be due to the very short effective sizes (4 and 2). This indicates the last two
iterations are very different from the previous ones. 

\begin{figure*}[htpb]
    \centering 
    \includegraphics[width=0.48\textwidth]{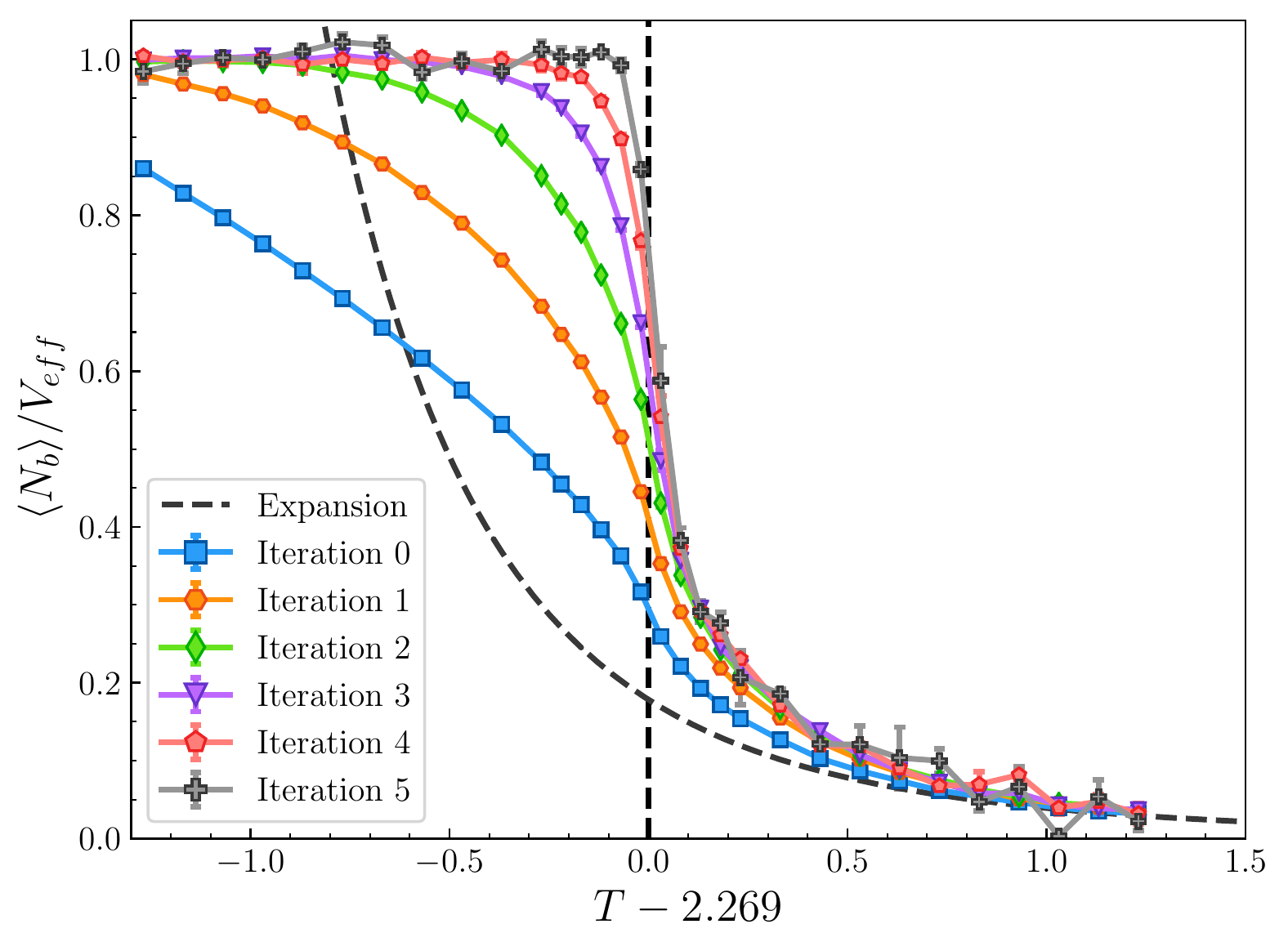}\hfill%
    \includegraphics[width=0.48\textwidth]{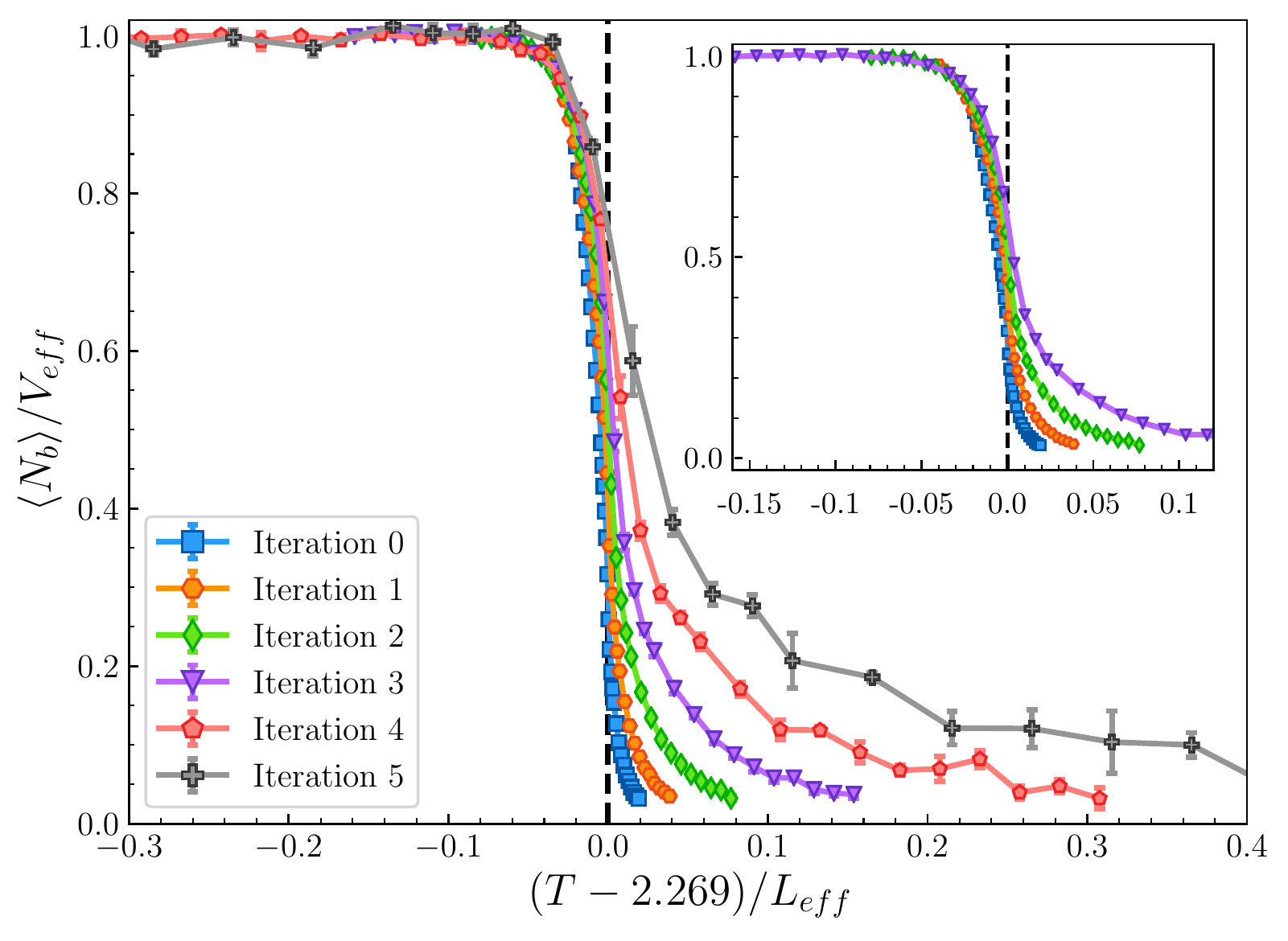}
    \caption{Average number of bonds $\langle N_b\rangle$ vs. temperature $T$
        under iterated blocking steps beginning with an initial lattice size of
        $L = 64$ (left). The dashed black line illustrates the high temperature
        expansion, showing that the dominant configurations are those
        consisting of small, isolated plaquettes.  Average number of bonds
        $\langle N_b\rangle$ vs. the rescaled temperature $(T - 2.269) /
        L_{eff}$ under successive blocking steps (right).  Iteration 0 represents the
    original lattice before blocking, with $L_{eff} = 64$. }
    \label{fig:imagecollapsed}
\end{figure*}

We now consider $\langle N_b\rangle$  for successive iterations. 
The results are shown in Fig. \ref{fig:imagecollapsed}. We see that in the low temperature side, the curves sharpen in a way 
similar to $T_{1100}$ in Fig. \ref{fig:T1100}. However on the high temperature side, we observe a merging rather than a crossing. 
This can be explained as follows. 
In the high T regime occasionally a single loop, the size of a plaquette, forms.  This is due to other configurations being highly suppressed. With the $1+1\rightarrow0$ rule,
one out of four possible plaquettes becomes a larger plaquette which exactly compensates the
change in $V_{eff}$ which is also reduced by a factor of four.  
%In the blocking procedure represented in
%Fig.~\ref{fig:110}, 
There are four kinds of plaquettes (see Fig.~\ref{fig:110}): those inside the blocks 
(they disappear after blocking), those between two neighboring blocks in the
vertical or horizontal direction (these are double links between the blocks and
so they disappear with the $1+1\rightarrow0$ rule), and finally those which 
share a corner with four blocks (they generate a larger plaquette); this type
can be seen at (4, 12) in Fig.~\ref{fig:110}.
% illustrations can be found near (4,12) in Fig.~\ref{fig:110}.

We now attempt to obtain data collapse 
for $\langle N_b \rangle/V_{eff}$ by performing a rescaling of the temperature axis with respect to the critical value as in Fig. \ref{fig:T1100}. After this
rescaling by a factor 2 at each iteration, we observe a reasonable collapse on the low-temperature side.  On the high
temperature side, since the unrescaled curves merge, the rescaling splits
them and there is no collapse on that side. This is illustrated in the bottom part of Fig. \ref{fig:imagecollapsed}.  

\section{TRG calculation of \texorpdfstring{$\langle N_b \rangle$}{<Nb>}}
\label{sec:nbtrg}

Using the tensor method we were able to calculate $\langle N_{b} \rangle$ to
compare with the worm algorithm.  Consider the equation for $\langle N_{b}
\rangle$ with $N_{b} = \sum_{l} n_{l}$ the sum over bond numbers at every link.

\begin{align}
    \langle N_{b} \rangle = \frac{1}{Z} \sum_{ \{ n \} } &\left( \sum_{l} n_{l}
    \right) \left( \prod_{l} \tanh^{n_{l}}(\beta) \right) \\ 
    \nonumber &\times \left( \prod_{i} \delta^{(i)}_{n_{x}+n_{x'}+n_{y}+n_{y'},
    0\text{ mod }2} \right).
\end{align}

This expression can be seen as $\langle N_{b} \rangle = \sum_{l} \langle n_{l}
\rangle$, and because of translation and $90^{\circ}$ rotational invariance,
all $\langle n_{l} \rangle$ are equal.  Thus, it is enough to calculate
$\langle n_{l} \rangle$ for one particular link (just call it $\langle
n\rangle$) and multiply it by $2V$: $\langle N_{b} \rangle = 2V \langle n
\rangle$.

To calculate $\langle n \rangle$, it amounts to associating an $n$ with one
particular link on the lattice.  This alters \emph{two} tensors on the lattice
such that the two tensors which contain that link as indices are now defined as
\begin{align}
	\tilde{T}^{(1)}_{n_{x} n_{x'} n_{y} n_{y'}}  &=
    \sqrt{n_{x}} T_{n_{x} n_{x'} n_{y} n_{y'}} (\beta)\\
    \tilde{T}^{(2)}_{n_{x} n_{x'} n_{y} n_{y'}}  &=
    \sqrt{n_{x'}}T_{n_{x} n_{x'} n_{y} n_{y'}} (\beta)
\end{align}
where $x$ and $x'$ were chosen without loss of generality.  It could just as
well have been chosen as $y$ and $y'$.  One can see that when these two tensors
are contracted along their shared link, the product picks up a factor of $n$
for that link, which when combined with the other tensors in the lattice, and
divided by $Z$, yields $\langle n \rangle$.

\begin{figure}[htpb]
    \centering
	\includegraphics[width=0.4\textwidth]{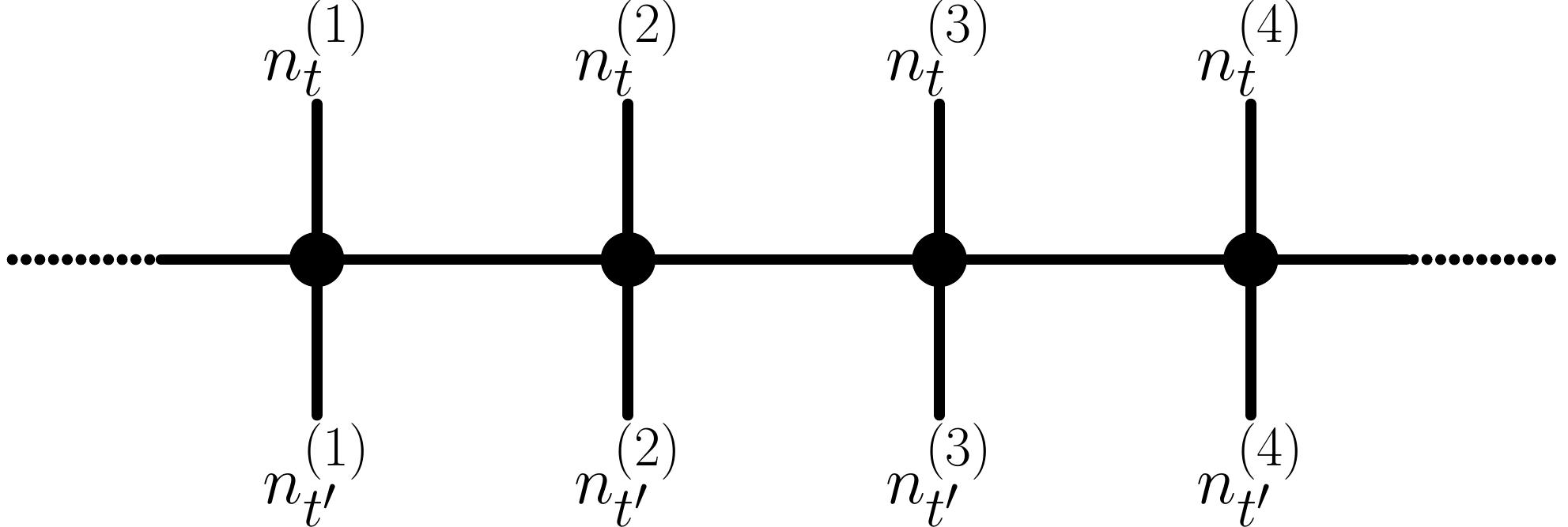}
    \caption{A pictorial representation of the transfer matrix made by
    contracting a fundamental tensor along a single time-slice.}
    \label{fig:tm}
\end{figure}

Knowing the above, one is free to block and construct the partition function,
$Z$, and $\langle n \rangle$.  This can be done by blocking symmetrically in
both directions, or by constructing a transfer matrix by contracting only along
a time-slice.  This is shown in Fig.~\ref{fig:tm}.  In practice contracting to
build a transfer matrix is optimum since one direction of the lattice is never
renormalized and allows the easy calculation of $\langle n \rangle$.
What is described in the previous section is a method to calculate $\langle n
\rangle$ for the original, fine, lattice.  However, one can also calculate the
same quantity for a coarse lattice.  The actual blocking method is essentially
identical, with a small exception.  Instead of contracting the fundamental
tensor to the desired lattice size, one contracts a blocked tensor to the
desired lattice size.

For example, if one wanted to calculate $\langle N_{b} \rangle$ for a $32
\times 32$ lattice, one could contract the fundamental tensor along a time
slice with itself five times.  This would give a $2^{32} \times 2^{32}$
transfer matrix which could be used to build the whole partition function.
Now, under a single coarse-graining step the $32\times 32$ lattice becomes a
$16\times 16$ lattice of blocks.  Therefore, to build this, one could contract
four fundamental tensors in a block and consider this a new, effective
fundamental tensor.  This is shown in Fig.~\ref{fig:unit_block}.
Then one repeats the same steps to construct the transfer matrix, however only
contracting four times with itself to create a matrix representing 16 lattice
sites of the blocked tensor.

\begin{figure}[htpb]
	\includegraphics[width=0.4\textwidth]{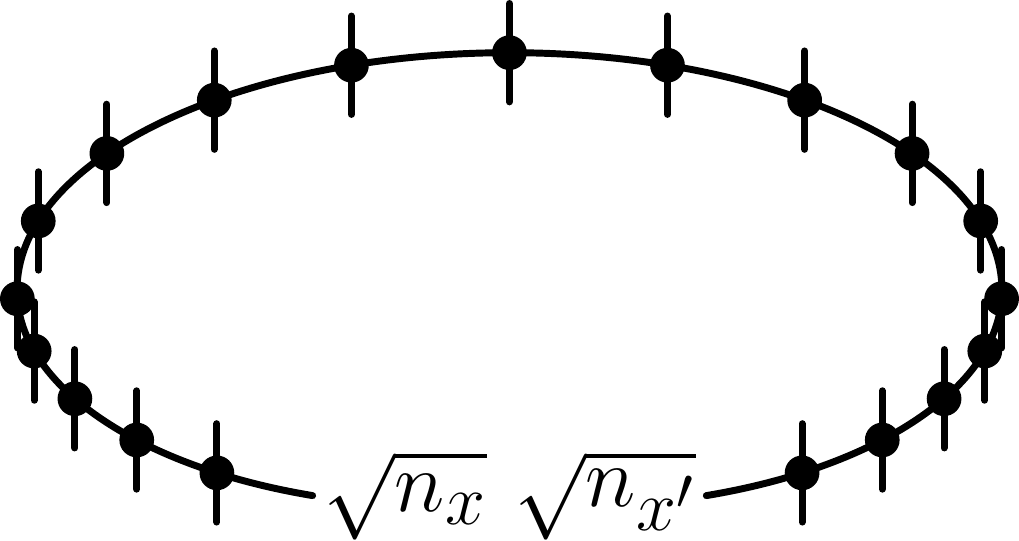}
    \caption{By multiplying the remaining free spatial indices by $\sqrt{n}$
    and contracting for periodic boundary conditions in space we form an
    ``impure'' transfer matrix.  Combining the resultant matrix with the
    original transfer matrix allows one to calculate $\langle n \rangle$.}
    \label{fig:nb}
\end{figure}

To actually calculate $\langle n \rangle$ by building the transfer matrix, one
can take the final tensor, prior to contracting the dangling spatial indices,
and multiply by $\sqrt{n}$ against the indices $n_{x}$ and $n_{x'}$. This is
shown for the unblocked case in Fig.~\ref{fig:nb}, however the procedure is
identical for the blocked case once the blocked tensor has been constructed.
This is also the point where one can choose the level of approximation one will
use in the blocking.  For instance one could choose that the state $|1 \; 1
\rangle \rightarrow |0\rangle$ and assign $n = 0$ to that state.  Alternatively
one could preserve $N_{b}$ and let $|1 \; 1 \rangle \rightarrow |2 \rangle$ and
assign $n = 2$ to that state.  This procedure was found to agree with the
results obtained by changing the pixels of the worm configurations.

\begin{figure*}[htpb]
    \centering 
    \includegraphics[width=0.48\textwidth]{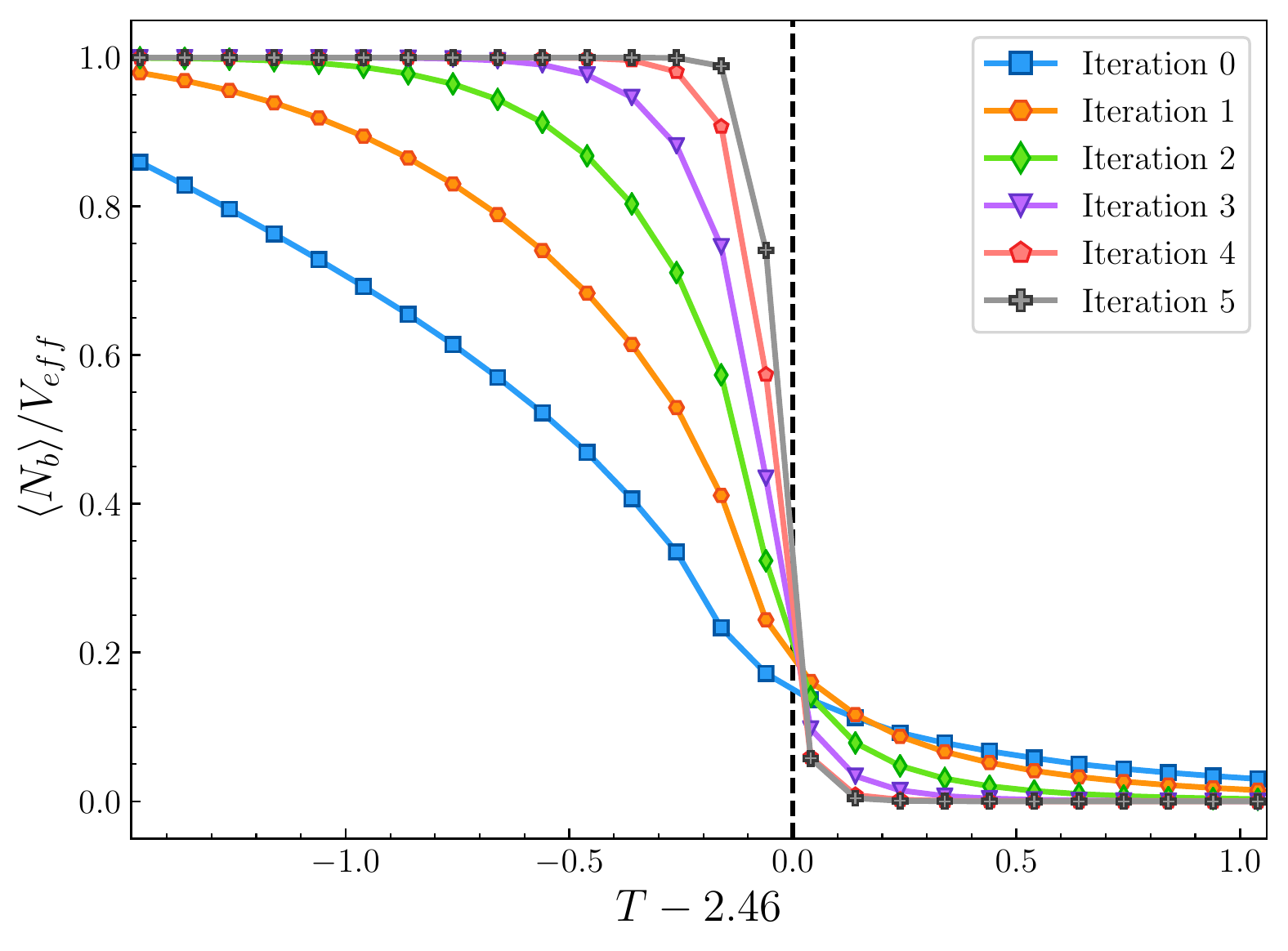}\hfill%
    \includegraphics[width=0.48\textwidth]{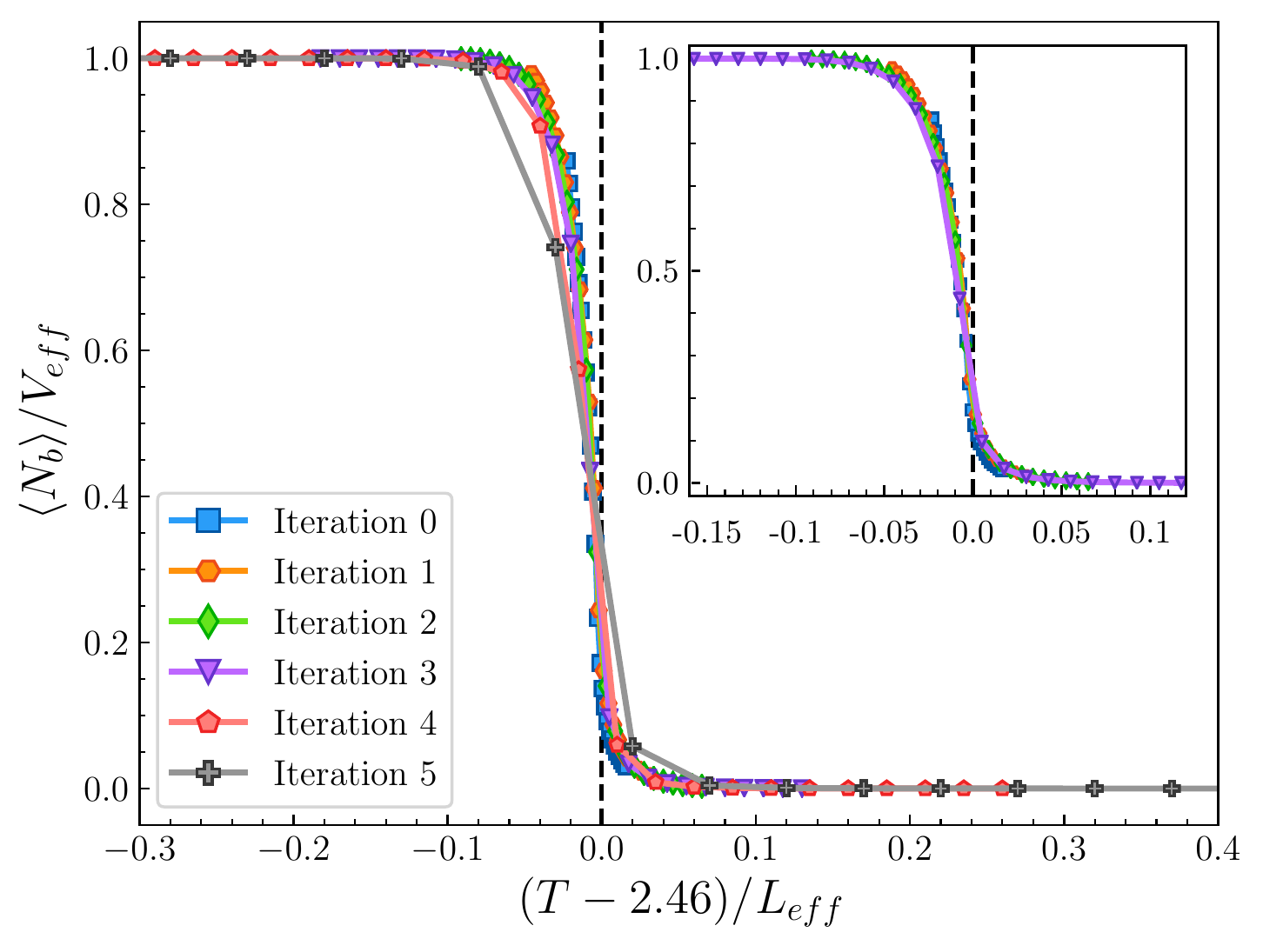}
    \caption{$\langle N_b\rangle$ vs $(T - 2.46)$ under successive blocking
        steps calculated using 2-state HOTRG (left). $\langle N_b\rangle$ vs 
        $(T - 2.46) / L_{eff}$ under successive blocking steps calculated using
    2-state HOTRG (right).}
    \label{fig:Nb_2s_HOTRG}
\end{figure*}

Once the original transfer matrix has been constructed, as well as the matrix
with the insertion of $n$ along a single link, one can combine these to find
$\langle n \rangle$.  This is done by simple matrix multiplication:
\begin{equation}
    \langle n \rangle = \frac{1}{Z} \tr[\underbrace{\mathbb{T} \cdots
    \mathbb{T}' \cdots \mathbb{T}}_{N_{\tau}}]
\end{equation}
with
\begin{equation}
	Z = \tr[\mathbb{T}^{N_{\tau}}].
\end{equation}
Here $\mathbb{T}$ represents the transfer matrix built by contracting tensors
along a time slice, and $\mathbb{T}'$ represents the single (``impure'')
transfer matrix at a time-slice with a single bond multiplied by $n$.  Since
the lattice has Euclidean temporal extent, $N_{\tau}$, there are that many
matrices multiplied in each case. The values of 
$\langle N_b\rangle/V_{eff}$ obtained with this procedure are shown in Fig. \ref{fig:Nb_2s_HOTRG}. 
The rescaling by 2 at each iteration provides 
a good data collapse on both sides of the transition.

\section{Conclusions}
In summary, we have motivated, constructed and applied a RG transformation to sets of worm configurations at various temperature. This transformation 
is approximate and the coarse-grained configurations are themselves worm configurations. This allows multiple iterations. We monitored the bond density at successive iterations and compared them with a two-state TRG 
approximation. We found clear similarities in the low-temperature side, 
where data collapse is observed for both procedures when the distance to 
the critical point is rescaled at each iteration. In the high-temperature phase, only the TRG approximation shows good data collapse. 

Can the procedure developed here be applied to the boundary of 
arbitrary sets of images as illustrated in the introduction?
The gray cutoff could be used as a tunable parameter. However, 
in the limiting cases of a zero (one) gray cutoff we have uniform 
black (white) images which are both similar to the high-temperature phase, and we do not expect a phase transition. 
Applications to the CIFAR dataset are discussed in Appendix \ref{sec:cifar} 
and confirm this point of view. 

A better understanding of RG concepts in machine learning could enhance physics discovery, especially
in the context of simulation and modeling of physical systems at a fundamental
level \cite{Shanahan:2018vcv}.
The general idea is to render computational tools ``smart,'' i.e.,
that they would learn features and patterns without the intervention of a human
``assistant,'' and would, in the best possible scenario, guide the direction of
further simulations. This could accelerate and deepen the process
of understanding and characterizing the complex systems that are deemed
important in pure and applied physics.

\section*{Acknowledgements}
This research was supported in part by the Department of Energy, Office of Science, Office
of High Energy Physics, Grant No.~DE-SC0013496 (JG)
 DE-SC0010113 (YM) and DE-SC0009998 (JUY) and 
 Office of Workforce Development for Teachers and Scientists, Office of Science Graduate Student Research 
(SCGSR) program. The SCGSR program is administered by the Oak Ridge Institute for Science and Education 
(ORISE) for the DOE. ORISE is managed by ORAU under contract number DE‐SC0014664 (SF).

\bibliography{centralmacbib2.bib,preising}

\clearpage

\appendix
\section{Technical results}
\label{sec:technical_results}
\subsection{Loop representation}   
\label{ssec:loop_representation}
We can rewrite the Ising partition function in terms of bonds between
neighboring sites $\langle i , j \rangle$. The allowed bond configurations
are concisely described by concepts in graph theory, because they form edges
(bonds) between neighboring vertices (sites). Making use of well-known
identities allows for the partition function to be written in the following
high-temperature expansion:
\begin{align}
    Z &= 2^{|V|} \cosh^{|E|} \beta \sum_{\Gamma \in {\cal C}(G)}
    \tanh^{|\Gamma|} \beta \\
    &= 2^{|V|} \cosh^{|E|} \beta \sum_{|\Gamma|} \tanh^{|\Gamma|} \beta ~
    n(|\Gamma|)
\label{eq:z}
\end{align}
The notation is as follows. We have a graph $G=(V,E)$ that describes our
lattice, where $V$ are the vertices and $E$ are the edges, which are the bonds
between neighboring sites. If we restrict ourselves to subgraphs with only
occupied bonds allowed by the Ising model, then the degree of each vertex is
even.  This is the number of bonds emanating from a particular vertex. The set
of edges of such a subgraph is described as being ``Eulerian.''  The space of
all such sets of edges is known as the cycle space ${\cal C}(G)$. The notation
$|V|$, $|E|$, $|\Gamma|$ denotes the number of elements in each set
(cardinality).  In the second line, $n(|\Gamma|)$ counts the multiplicity of
subgraphs of cardinality $|\Gamma|$, and is zero when $|\Gamma|$ does not
correspond to a ``legal'' subgraph.

We now specialize the presentation to the the case of the two-dimensional Ising
model on a square lattice with periodic boundary conditions. In this case $|V|$
is $V=L^2$, the volume that we express in lattice units, and $|E| = 2V$. We
introduce the notation $t\equiv\tanh(\beta)$ and we call $N_b$ the number of
bonds in a graph (values taken by $|\Gamma|$). With these notations
we recover Eq.~\ref{eq:nbsum}.

It is this bond formulation that is the basis of both random cluster algorithms
\cite{Swendsen:1987ce} and worm algorithms \cite{prok87}.  In this paper we
utilize the latter.  Both of these classes of algorithms have the benefit of
significantly avoiding critical slowing down.  This is essential near the
critical temperature $T_c$.
\subsection{Heat capacity}
\label{ssec:heat_capacity}
One striking feature of the second order transition for the two-dimensional
Ising model is the logarithmic divergence of the specific heat density at the
critical temperature $T_c$. In this section, we review the way the specific
heat can be calculated with the worm algorithm and we check our answer with the
exact finite volume expressions~\cite{bkaufman}.

Using the standard thermodynamical formula for the average energy
\begin{equation}
    \langle E \rangle = - \frac{\partial}{\partial \beta} \ln Z,
\end{equation}
with the expression (\ref{eq:nbsum}) of  $Z$, we get
\begin{equation}
    \langle E \rangle = - \tanh(\beta)\left(2V + \frac{\langle
    N_b\rangle}{\sinh^2(\beta)}\right),
\end{equation}
where we define
\begin{equation}
\langle f(N_b)\rangle \equiv \sum_{N_b} f(N_b)t^{N_b}  n(N_b)/\sum_{N_b}t^{N_b}
n(N_b)
\end{equation}
We can then use
\begin{equation}
    C_{V} = \frac{\partial \langle E \rangle}{\partial T},
\end{equation}
to write
\begin{align}
    \frac{C_V}{V} = \beta^2 \bigg[&\frac{2}{\cosh^2(\beta)} -
        \frac{4\cosh(2\beta)}{\sinh(2\beta)} \frac{\langle N_b\rangle}{V} \\
    &+ \left(\frac{2}{\sinh(2\beta)}\right)^2\frac{\langle(N_b - \langle
           N_b\rangle)^2\rangle}{V} \bigg]
\end{align}

%Singularity of $C_V$ at $T = T_C$
Since $\frac{\langle N_b\rangle}{V}\leq 2$ (in two dimensions), the only
possibly divergent part is the variance of $N_b$ per unit volume $\langle
\Delta_{N_b}^2\rangle$ defined in Eq.~(\ref{eq:fluc}). The singularity near
$T_c$ is known from Onsager's solution:
\begin{equation}
    \frac{C_V}{V} = -\frac{2}{\pi}\left(\ln(1+\sqrt{2})\right)^2
        \ln\left(|T - T_c|\right) + \text{regular}.
    \label{eq:onsager}
\end{equation}
This implies Eq.~(\ref{eq:specific_heat_fluctuation_eq2}).
\subsection{Monte Carlo implementation}
\label{ssec:monte_carlo_implementation}
We can proceed to sample the closed path configuration space using the worm
algorithm \cite{prok87}. A single Monte Carlo step is outlined below.
\begin{enumerate}
    \item Randomly select a starting point on the lattice $(i_0, j_0)$.
    \item Propose a move to a neighboring site $(i^{\prime}, j^{\prime})$,
        selected at random.
    \item If no link is present between these two sites, a bond is created with
        acceptance probability $P = \min\{1, \tanh{\beta}\}$. If the bond is
        accepted, we update the bond number for the present worm, $n_b = n_b +
        1$.
    \item If a link already exists between the two sites, it is removed with
        probability $P = 1$.
    \item If $(i^{\prime}, j^{\prime}) = (i_0, j_0)$, i.e. we have a closed
        path, go to (1.). Otherwise, $(i^{\prime}, j^{\prime}) \neq (i_0,
        j_0)$, go to (2.)
\end{enumerate}
The number of necessary Monte Carlo steps required to achieve sufficient
statistics varies with the lattice size, thermalization time, and temperature.
After each step, we calculate the energy in terms of the average number of
active bonds $N_b$, and consider the system to be thermalized when
fluctuations between subsequent values of the energy are less than
$1\times10^{-3}$. We then save the resulting configuration, along with the final
values for all physical quantities of interest. This process is then repeated
many times over a range of different temperatures to generate sufficient
statistics for physical observables. All errorbars are calculated using the
block jackknife resampling technique. 
\subsection{Tests}
\label{ssec:tests}
\begin{figure}[htpb]
    \centering
    \includegraphics[width=8.6cm]{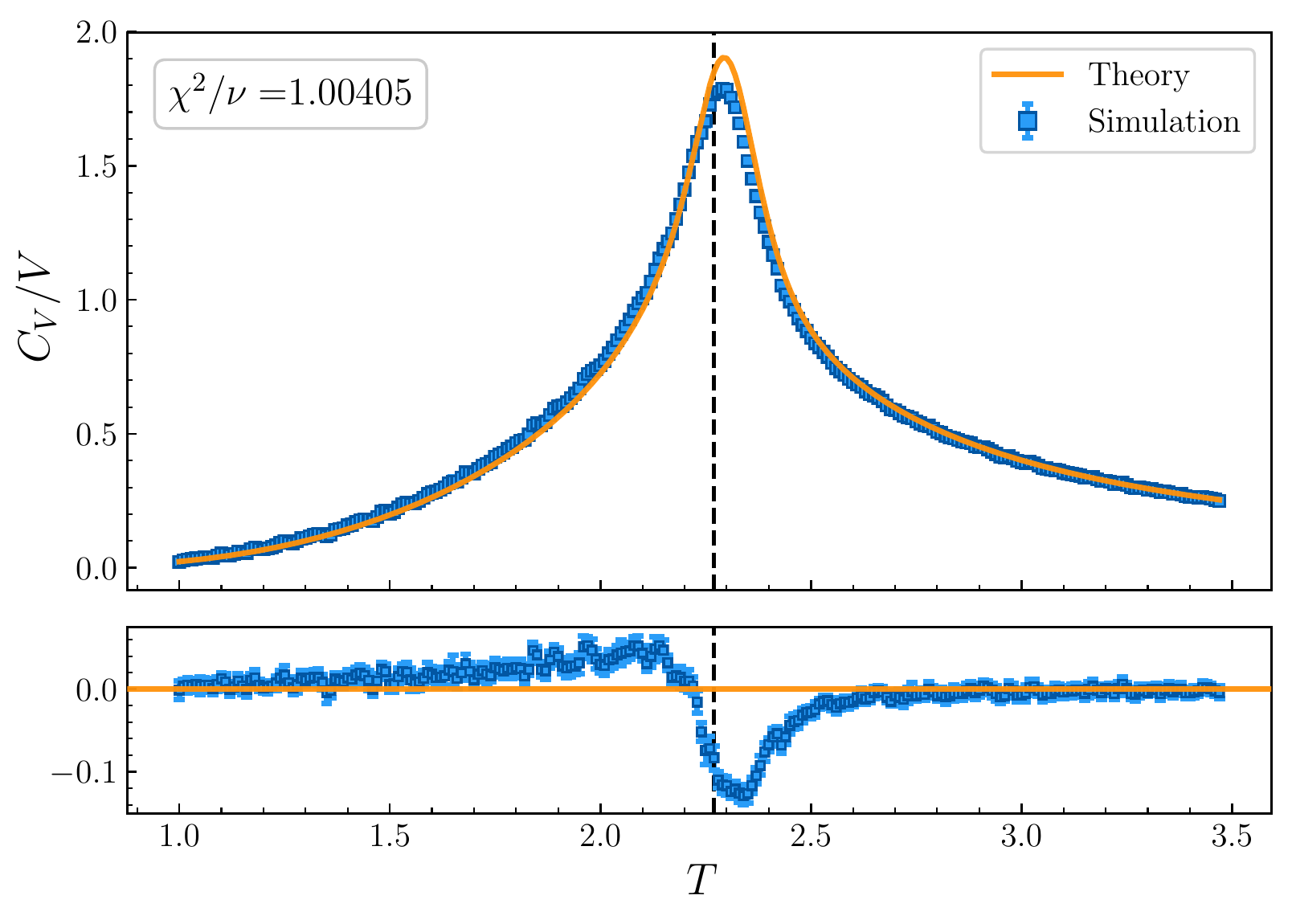}
    \caption{Comparison of the worm Monte Carlo computation of the specific
        heat $C_v$ versus temperature and the exact results using the formula
        in \cite{bkaufman}, for an $L=32$ lattice. It can be seen that the
        agreement is excellent, except for a slight deviation at the critical
        temperature, where Monte Carlo algorithms tend to face difficulties
        with critical slowing down.  This is mostly addressed with the worm
        algorithm, in terms of having a dynamical scaling exponent that is zero
        rather than two, but there is (as can be seen), still a residual
        suppression of fluctuations in the immediate vicinity of $T_c$.
    \label{Kcompare}}
\end{figure}

The above formulas have been used to calculate $C_V$.  Precise checks were
performed by comparing with with the exact results obtained from
Ref.~\cite{bkaufman}. The agreement  can be seen in Fig.~\ref{Kcompare}.
Results for other lattice sizes that we have simulated are similar.

\subsection{Conjecture about \texorpdfstring{$\lambda_{max}$}{λmax}}
\label{ssec:conjecture}
\begin{figure}[htpb]
    \centering
    \includegraphics[width=8.6cm]{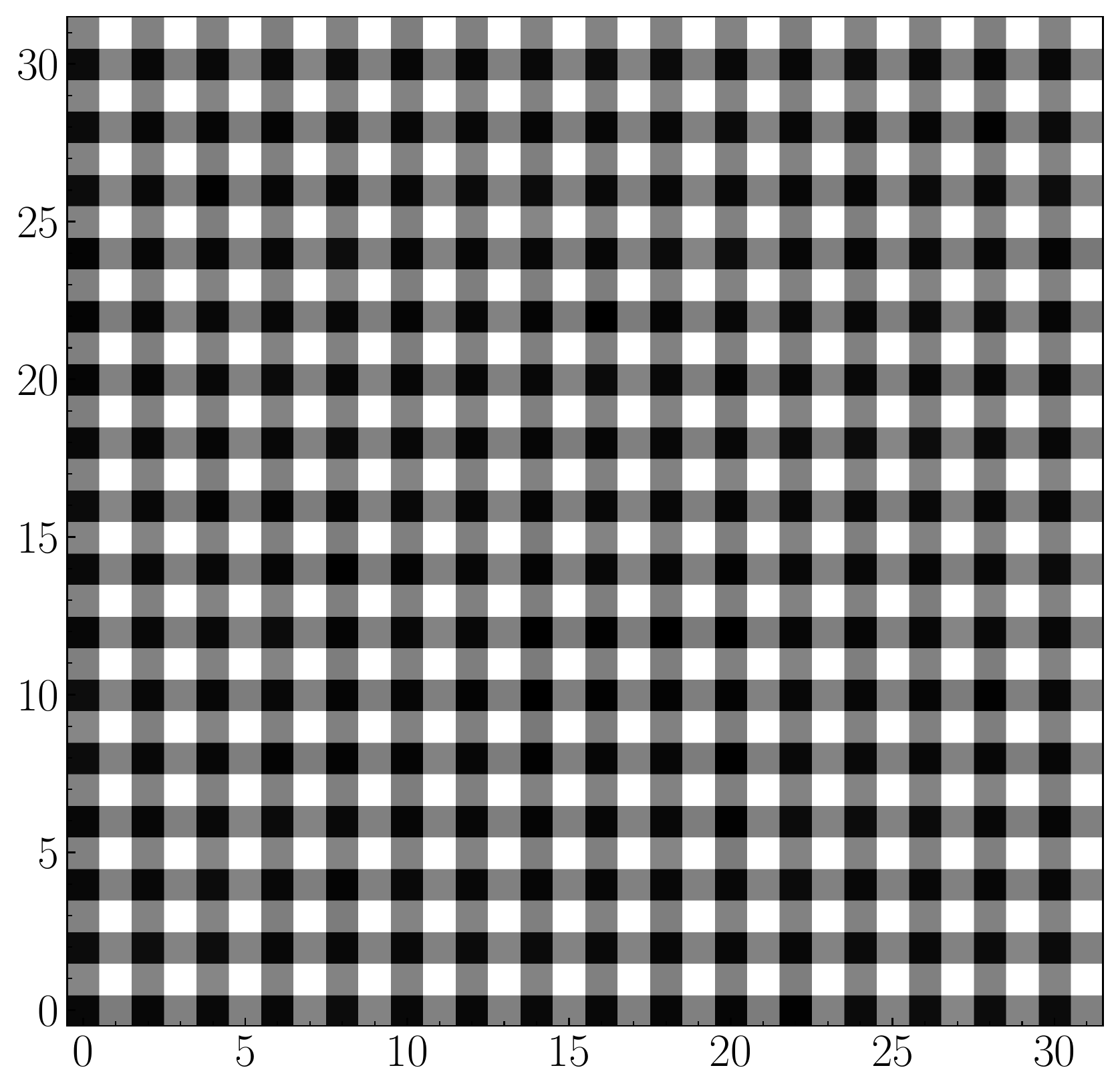}
    \caption{Average image $\langle \mathbf{v}\rangle$ calculated for the
        $L = 16$ lattice at $T = 2.0$, illustrating the
        tablecloth-like appearance.}
    \label{fig:average_image}
\end{figure}

Using the Monte Carlo algorithm outlined above, we can calculate the average number of
occupied bonds at a particular temperature by averaging over all configurations
% \begin{equation}
%     \langle \mathbf{v}\rangle = \frac{1}{N_{configs}}\sum_{\{configs\}}
%     \mathbf{v}
%     \label{eq:pixel_avg}
% \end{equation}
\begin{align}
    % \label{eq:pixel_avg}
    % \langle \mathbf{v}\rangle
    %     &\equiv \frac{1}{N_{configs}}\sum_{\{configs\}} \mathbf{v}\\
    \langle N_b\rangle
        &\equiv \frac{1}{N_{configs}}\sum_{n=1}^{N_{configs}} N_b^{(n)}\\
        &= \left\langle \sum_{j=bonds} v_j \right\rangle \\
        &= 2V\langle \mathbf{v}_b\rangle  \Longrightarrow\\
        \langle \mathbf{v}_b\rangle &= \frac{\langle N_b\rangle}{2V},
    \label{eq:link_avg}
\end{align}

where we've defined $\langle \mathbf{v}_b\rangle$ to be the average occupation
of bonds, $N_b^{(n)}$ to be the number of occupied bonds in the $n$-th
configuration, and we've used (\ref{eq:link_sum}) in the second line. If we
consider graphs with no self-intersections,
\begin{equation}
    \sum_{j=bonds} v_j = \sum_{j=sites} v_j.
    \label{bonds_equal_sites}
\end{equation}
For small $\beta$ (high $T$) this can be a good approximation,
\begin{align}
    \left\langle \sum_{j=bonds} v_j \right\rangle &\simeq \left\langle
    \sum_{j=sites} v_j\right\rangle\Longrightarrow\\
    \langle\mathbf{v}_b\rangle &\simeq \frac{1}{2}\langle\mathbf{v}_s\rangle
    \label{avg_bonds_equal_sites}
\end{align}
This agrees with our intuition, that the average image $\langle
\mathbf{v}\rangle$ should resemble a ``tablecloth'', where the site pixels are
twice as dark as the link pixels. This can be seen clearly in
Fig.~\ref{fig:average_image}.

For a general graph, a link is shared by two sites (its endpoints), whereas a
site can be shared either by 0, 2, or 4 bonds. If the site is shared by two
bonds, it is only visited once, denoted $sites^{(1)}$, and if it is shared by
four bonds, it is visited twice, denoted $sites^{(2)}$. This allows us to
break up the sum over bonds into two terms
\begin{equation}
    \sum_{j=bonds} v_j = \frac{2}{2}\sum_{j=sites^{(1)}} v_j
        + \frac{4}{2} \sum_{j=sites^{(2)}} v_j
\end{equation}
Rearranging and taking averages gives
\begin{align}
    \left\langle \sum_{j=sites^{(1)}} v_j \right. 
    + &\left. \sum_{j=sites^{(2)}} v_j \right\rangle\\
    &=\left\langle \sum_{j=bonds} v_j - \sum_{j=sites^{(2)}} v_j \right\rangle\\
    &= \left\langle \sum_{j=sites} v_j\right\rangle\\
    &= V\left\langle \mathbf{v}_s\right\rangle\\
    & = \left\langle \sum_{j=bonds} v_j \right\rangle - \left\langle
        \sum_{j=sites^{(2)}} v_j \right\rangle\\
    &= 2V \langle\mathbf{v}_b\rangle - \left\langle
        N_{sites^{(2)}}\right\rangle\Longrightarrow\\
    \frac{\left\langle N_{sites^{(2)}}\right\rangle}{V} &=
        2\langle\mathbf{v}_b\rangle - \langle \mathbf{v}_s\rangle.
\end{align}
We can rewrite the last equation using (\ref{eq:link_avg})
\begin{equation}
    \langle \mathbf{v}_s \rangle= \frac{\langle N_b\rangle}{V} - \frac{\langle
        N_{sites^{(2)}}\rangle}{V}.
\end{equation}
This suggests that a departure from a perfect tablecloth ($\langle
\mathbf{v}_s\rangle = 2\langle \mathbf{v}_b\rangle$) contains information.
Another useful construct is the covariance matrix,
% \begin{equation}
%     C = (V - \langle V\rangle)^T(V - \langle V\rangle)
% \end{equation}
%
% with entries,
\begin{align}
    C_{ij} &=
    \left\langle\left(v_i -\langle \mathbf{v}\rangle_i\right)
    \left(v_j -\langle \mathbf{v}\rangle_j\right)^{T}\right\rangle\\
        &= \frac{1}{N_{configs}}\sum_{n=1}^{N_{configs}}
        \left(v_i^{(n)} -\langle \mathbf{v}\rangle_i\right)
        \left(v_j^{(n)} -\langle \mathbf{v}\rangle_j\right)^{T},
    \label{covariance_matrix}
\end{align}
where we've defined $v_k^{(n)}$ to be the grayscale value of the $k$-th pixel
in the $n$-th sample configuration, and $\langle \mathbf{v}\rangle_k$ is the
average grayscale value of the $k$-th pixel over the set of configurations.
% whose significance will become apparent in later sections.

At some fixed temperature, the covariance matrix, $C \in
\mathbb{R}^{N_{configs}\times4L^2}$, where $N_{configs}$ is the number of sample
configurations (images), with each configuration flattened into a vector of
length $4L^2$. We can then perform a singular value decomposition (SVD) on the
covariance matrix,
\begin{equation}
    C = W\Lambda W^{T}
    \label{svd}
\end{equation}
where $W$ is a $4L^2\times 4L^2$ matrix whose columns ($\mathbf{w}_k$) are the
eigenvectors of $C$, and $\Lambda$ is the diagonal matrix of the absolute value of the eigenvalues
$\lambda^{(k)}$ of $C$, arranged in decreasing order. Without loss of
generality, we can assume the eigenvectors $\mathbf{w}_k$ are real and
normalized such that $\mathbf{w}_k^{T} \mathbf{w}_k = 1$. Thus, we can write
\begin{align}
    C \mathbf{w}_k &= \lambda^{(k)} \mathbf{w}_k\\
    \mathbf{w}_k^{T} C \mathbf{w}_k &= \lambda^{(k)}
\end{align}
For our purposes, we are interested in the first principal component, with
eigenvalue $\lambda^{(1)} \equiv \lambda_1$ and corresponding eigenvector
$\mathbf{w}_1$.
% \equiv \mathbf{w}^{max}$.

We conjectured that the first principal component, $\mathbf{w}_1$ of the
covariance matrix $C$ is directly proportional to the average worm
configuration (image) $\langle \mathbf{v} \rangle$, i.e.
\begin{equation}
    \mathbf{w}_1 \propto \langle \mathbf{v}\rangle
    \label{conjecture}
\end{equation}

From our results in \ref{configs_as_images}, we can write
\begin{align}
    % \sum_j \langle \mathbf{v}_j \rangle \langle \mathbf{v}_j \rangle =
    \langle \mathbf{v}\rangle^2
    &= \langle \mathbf{v}\rangle^{T} \langle \mathbf{v}\rangle \\
    &= 2V\bondavg^2 + V \siteavg^2
    % 2V\langle \mathbf{v}_b \rangle^2 + V \langle \mathbf{v}_s\rangle^2
\end{align}
This suggests that
\begin{equation}
    \mathbf{w}_1 = \frac{\langle \mathbf{v}\rangle}{\sqrt{(2\bondavg^2 +
\siteavg^2)V}}
\end{equation}
Moreover, we can write
\begin{align}
    \sum_i \bigg(v_{i}^{(n)}&-\langle\mathbf{v}\rangle_i \bigg)
        \langle\mathbf{v}\rangle_i\\
        &= \bigg\{\bondavg\sum_{j=bonds}v_j^{(n)}+\siteavg\sum_{j=sites}v_j^{(n)}\\
        &\quad\quad\quad-\left(2V\bondavg^2+V \siteavg^2\right)\bigg\}\\
        &= \bigg\{\bondavg N_b^{(n)}+\siteavg N_{s}^{(n)}\\
        &\quad\quad\quad-V\left(2\bondavg^2+\siteavg^2\right)\bigg\}\\
        &= \bigg\{\bondavg\left(N_b^{(n)}-\langle N_b\rangle\right)\\
        &\quad\quad\quad+\siteavg \left(N_s^{(n)}-\langle N_s\rangle\right)\bigg\}\\
    &\equiv \bondavg \Delta_{N_b}^{(n)} + \siteavg \Delta_{N_s}^{(n)}
\end{align}
From this, we can extract a relationship between the eigenvalue corresponding
to the first principal component, $\lambda^{(1)}$ and the fluctuations
$\Delta_{N_b}$ and $\Delta_{N_s}$,
\begin{align*}
    \mathbf{w}_1^{T} C \mathbf{w}_1
        &= \lambda^{(1)}\\
        &= \frac{1}{N_{configs}}\sum_{n=1}^{N_{configs}} (
            \bondavg^2 \left(\Delta_{N_b}^{(n)}\right)^2   \\
            &+  2\bondavg \siteavg \Delta_{N_b}^{(n)} \Delta_{N_s}^{(n)}
            + \siteavg^2 \left(\Delta_{N_s}^{(n)}\right)^2
        )\\
        & \frac{1}{\left(2\bondavg^2 + \siteavg^2\right)}
\end{align*}

Now, if we consider the high temperature approximation where sites only have
single visits (no self-intersections), $\siteavg \simeq 2\bondavg$, $\langle
N_s\rangle \simeq \langle N_b\rangle$, and $\Delta_{N_b} \simeq \Delta_{N_s}$,
we have that $2\bondavg^2 + \siteavg^2 \simeq 6\bondavg^2$. and
\begin{align}
    \lambda_1 &\simeq \frac{\bondavg^2}{N_{configs}}\frac{9}{6\bondavg^2}
    \sum_{n=1}^{N_{configs}} \left(\Delta_{N_b}^{(n)}\right)^2 \\
    &= \frac{3}{2}\frac{1}{N_{configs}}\sum_{n=1}^{N_{configs}}
    \left(\Delta_{N_b}^{(n)}\right)^2 \\
    &= \frac{3}{2}\left \langle \Delta_{N_b}^2 \right\rangle.
    \label{eq:eigval_delta_Nb2}
\end{align}

\begin{figure}[htpb]
    \centering
    \includegraphics[width=8.6cm]{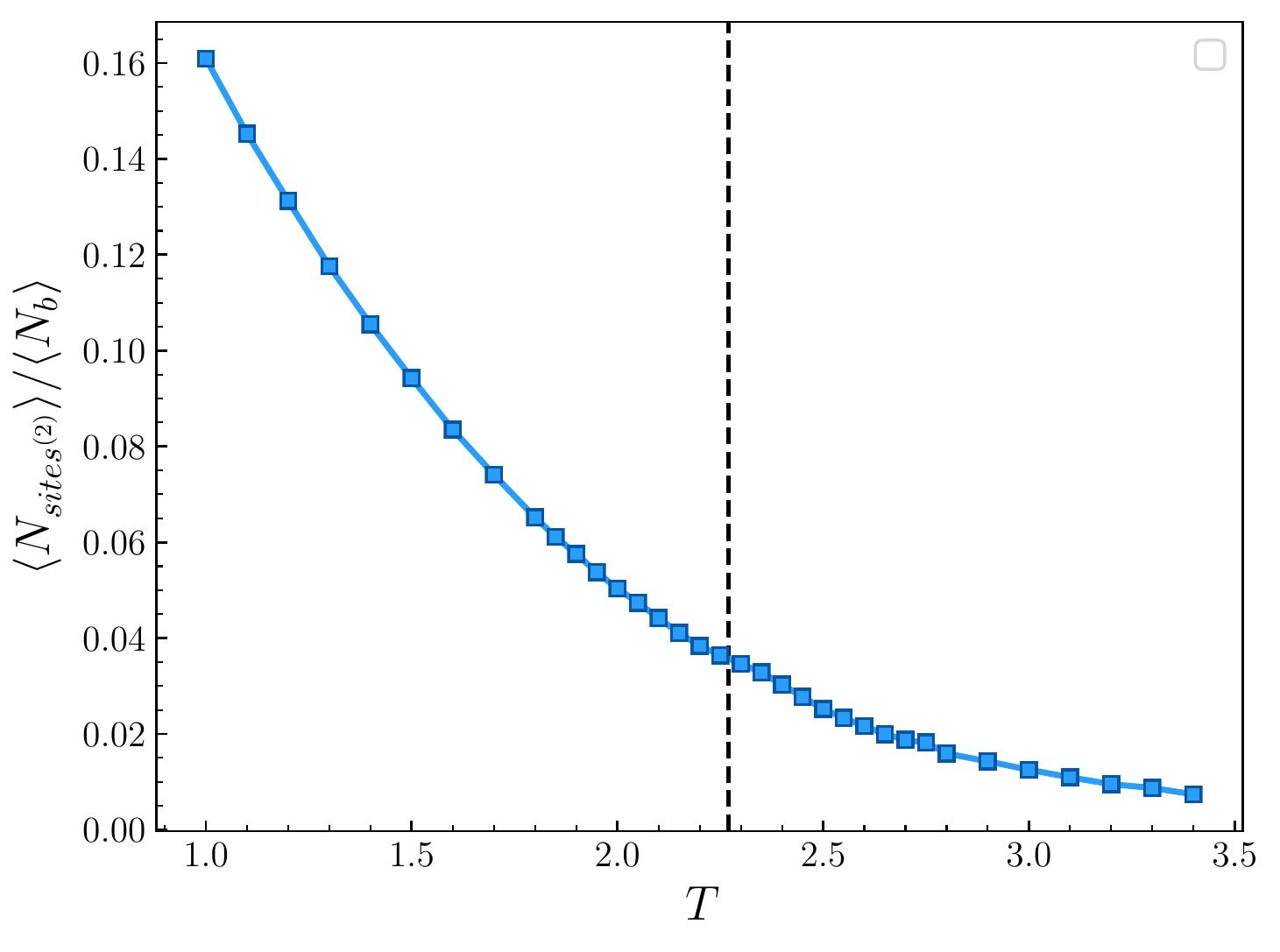}
    \caption{Ratio of the number of twice visited sites $\langle
    N_{sites^{2)}}\rangle$ to the average number of bonds $\langle N_b\rangle$
    versus temperature, for the $L=32$ lattice. This clearly justifies our
    approximation $\siteavg \simeq 2\bondavg$, where we ignore the contribution
from twice visited sites.}
    \label{fig:twice_visited_sites_ratio}
\end{figure}

A justification for making this approximation can be seen in
Fig.~\ref{fig:twice_visited_sites_ratio}. 
\subsection{Illustration of alternate blockings}
\label{subsec:altb}

\begin{widetext}

\begin{figure}[htpb]
    \centering
    %\subfigure[Unblocked]{
    %    \centering
    %    \includegraphics[width=0.31\textwidth]{configurations/worm_configuration_16_unblocked}
    %    \label{fig:unblocked}
    %    % \caption{Unblocked}
    %}\hfill
    %\subfigure[ $1 + 1 = 0$]{
    %    \centering
    %    \includegraphics[width=0.31\textwidth]{configurations/worm_configuration_16_blocked110}
    %    \label{fig:blocked_approx}
    %    % \caption{$1+1\rightarrow 0$}
    %}\hfill
    \subfigure[ $1 + 1 = m$]{
        \centering
        \includegraphics[width=0.31\textwidth]{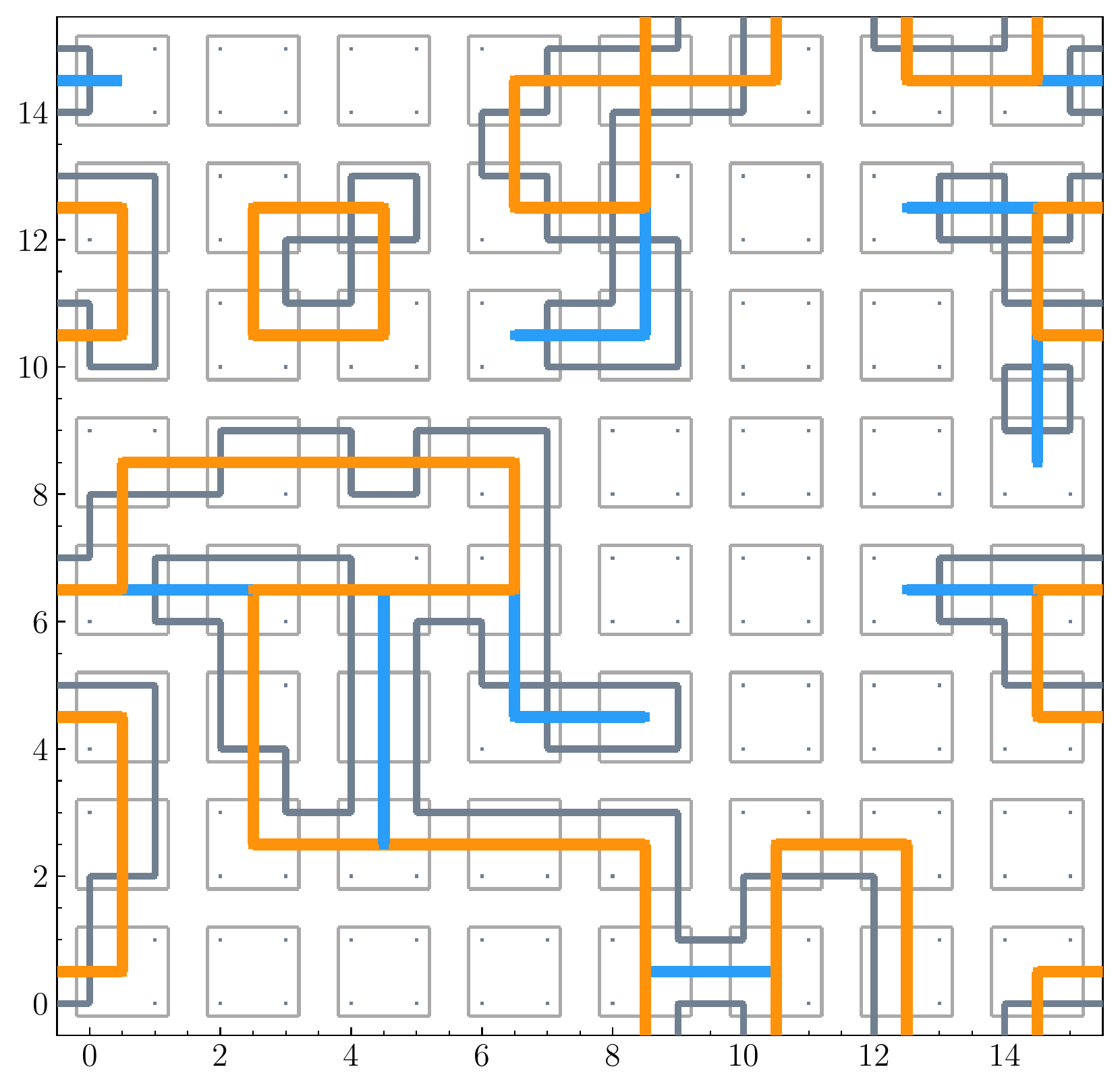}
        \label{fig:blocked_exact2}
        % \caption{$1+1\rightarrow 2$}
    }\hfill
% \end{figure}
% \begin{figure}[!h]
    \centering
    %\subfigure[ $1+1\rightarrow 0$]{
    %    \centering
    %    \includegraphics[width=0.31\textwidth]{configurations/worm_configuration_16_blocked110_image}
    %    \label{fig:blocked_approx_pca}
    %    % \caption{Unblocked}
    %}\hfill
    \subfigure[ $1+1\rightarrow 1$]{
        \centering
        \includegraphics[width=0.31\textwidth]{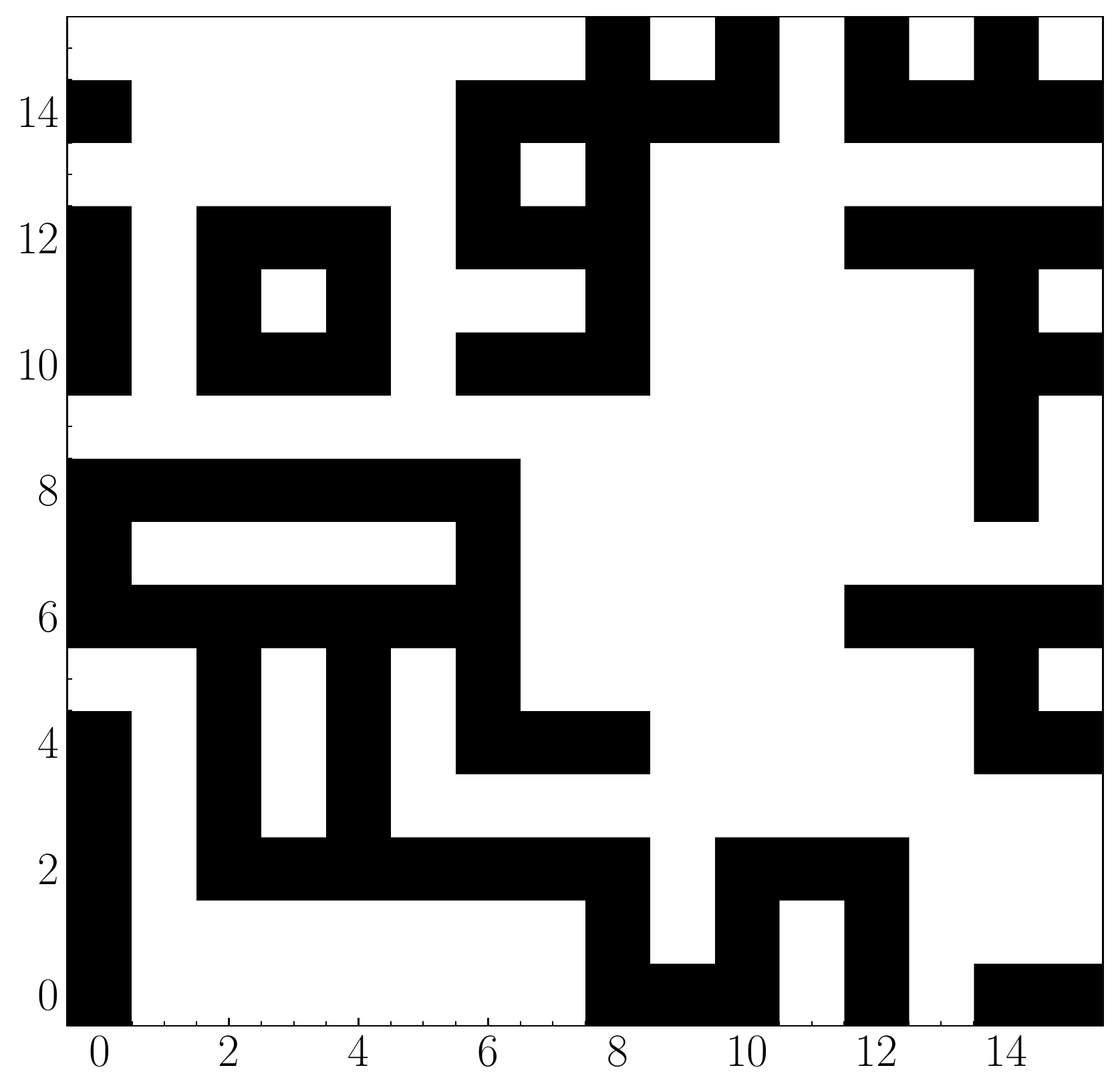}
        \label{fig:blocked_exact1_pca}
        % \caption{$1+1\rightarrow 0$}
    }\hfill
    \subfigure[ $1+1\rightarrow 2$]{
        \centering
        \includegraphics[width=0.31\textwidth]{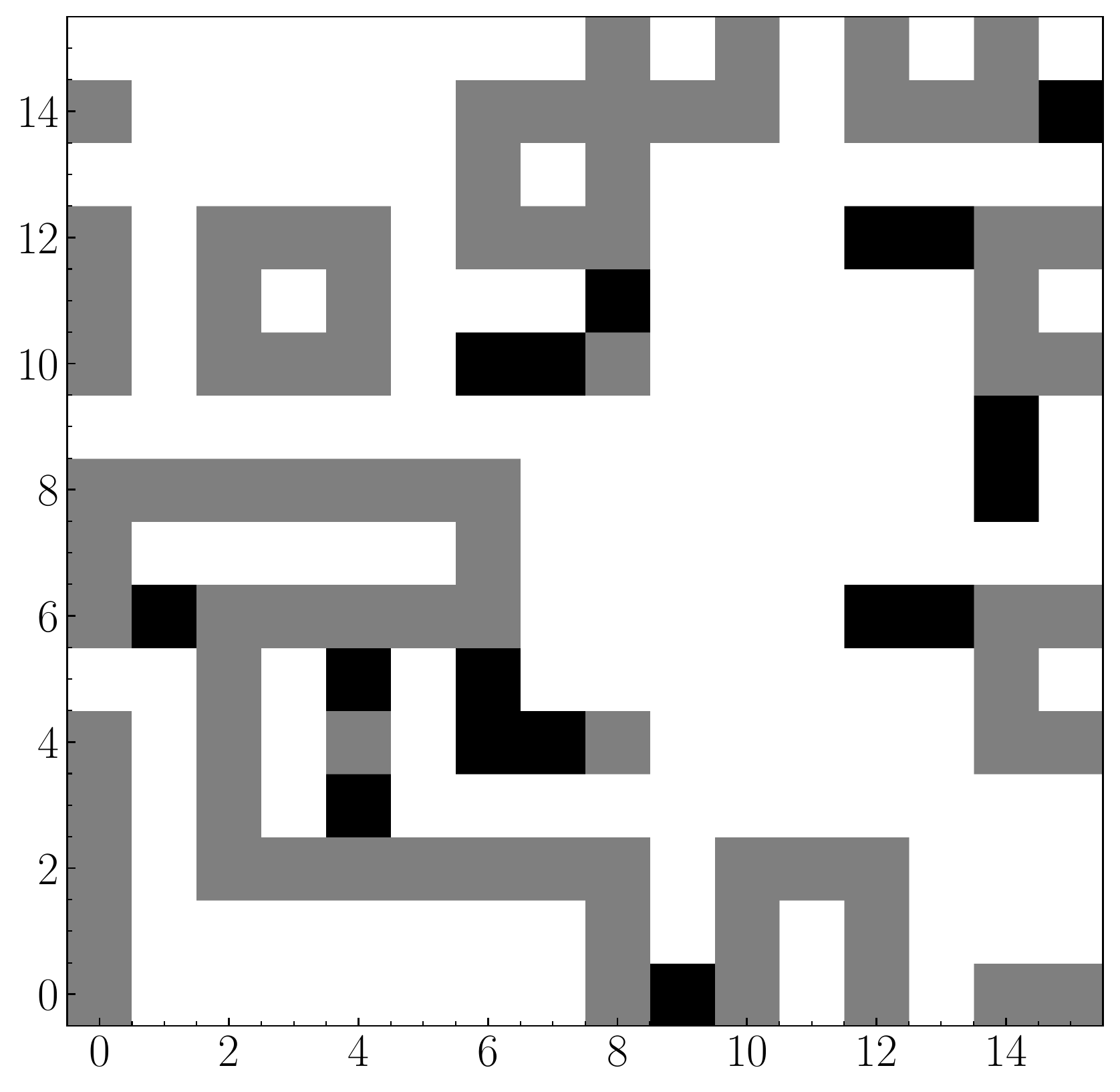}
        \label{fig:blocked_exact2_pca}
        % \caption{$1+1\rightarrow 2$}
    }\hfill
    \caption{Example of the different
        coarse-graining (``blocking'') procedures applied to a sample worm
        configuration generated at $T = 2.0$. Note that in
        (\ref{fig:blocked_exact2}) $m \in \{1, 2\}$, and double bonds are represented by blue lines.  (\ref{fig:blocked_exact1_pca}), (\ref{fig:blocked_exact2_pca}),
        illustrate the results of applying different weights to the so-called
        ``double bonds'' in the images representing a blocked configuration. Note that in (\ref{fig:blocked_exact1_pca}) $1+1\rightarrow 1$, double bonds are
        given the same weight as single bonds, and in
        (\ref{fig:blocked_exact2_pca}) $1+1\rightarrow 2$, double bonds are given twice the
        weight as single bonds, appearing twice as
        dark.}\label{fig:double_bond_weights}
    % \label{fig:blocked_config}
    % \caption{
\end{figure}
\end{widetext}
\section{Possible applications: From Images to Loops}
\label{sec:cifar}

\begin{figure*}[htb]
    \centering
    \includegraphics[width=0.31\textwidth]{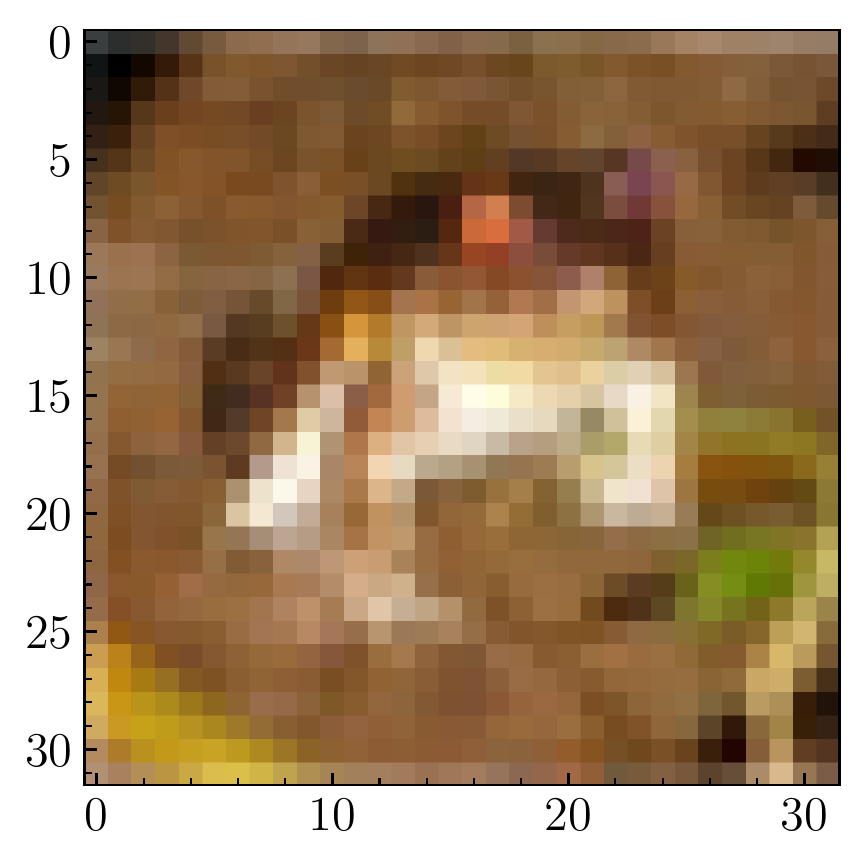}\hspace{2cm}%
    \includegraphics[width=0.31\textwidth]{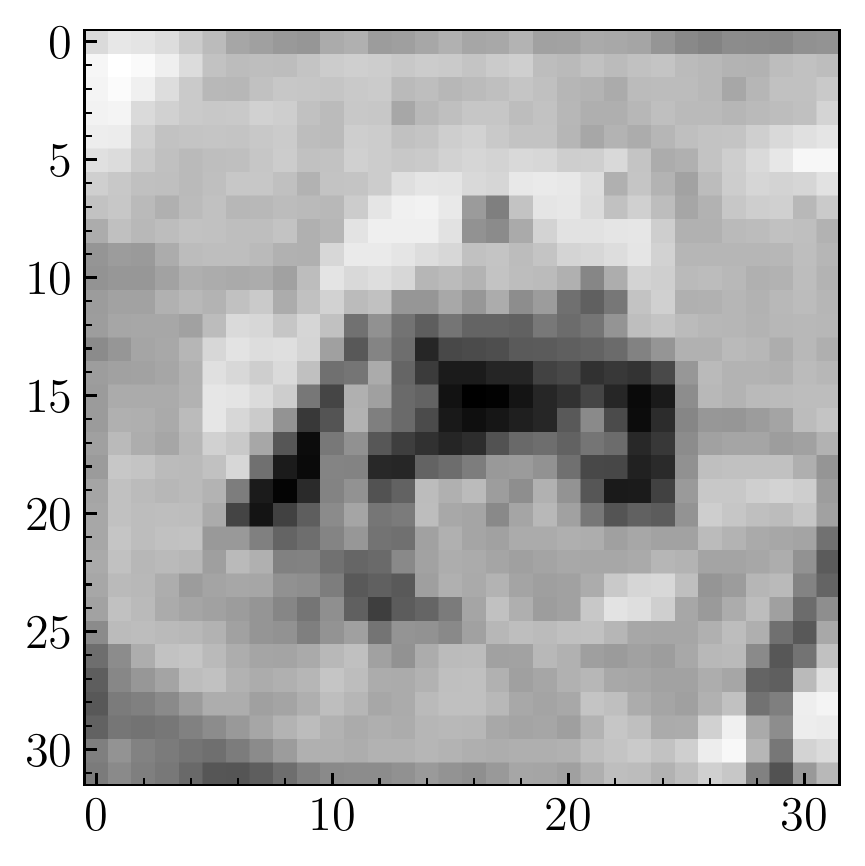}
    \includegraphics[width=0.31\textwidth]{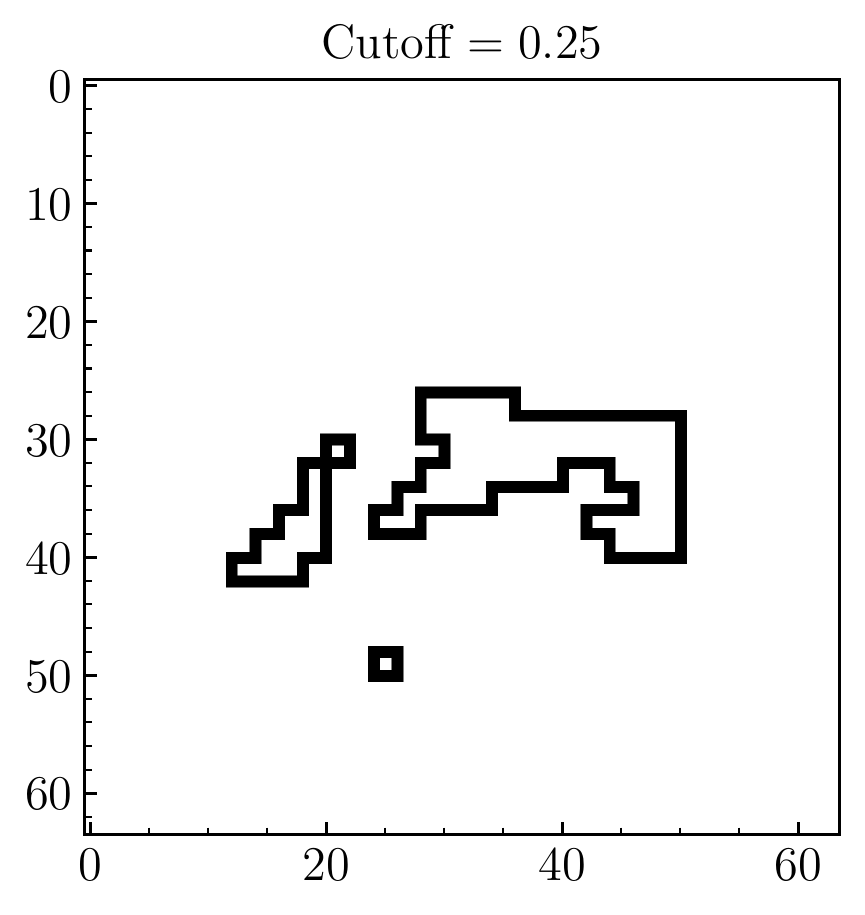}\hfill
    \includegraphics[width=0.31\textwidth]{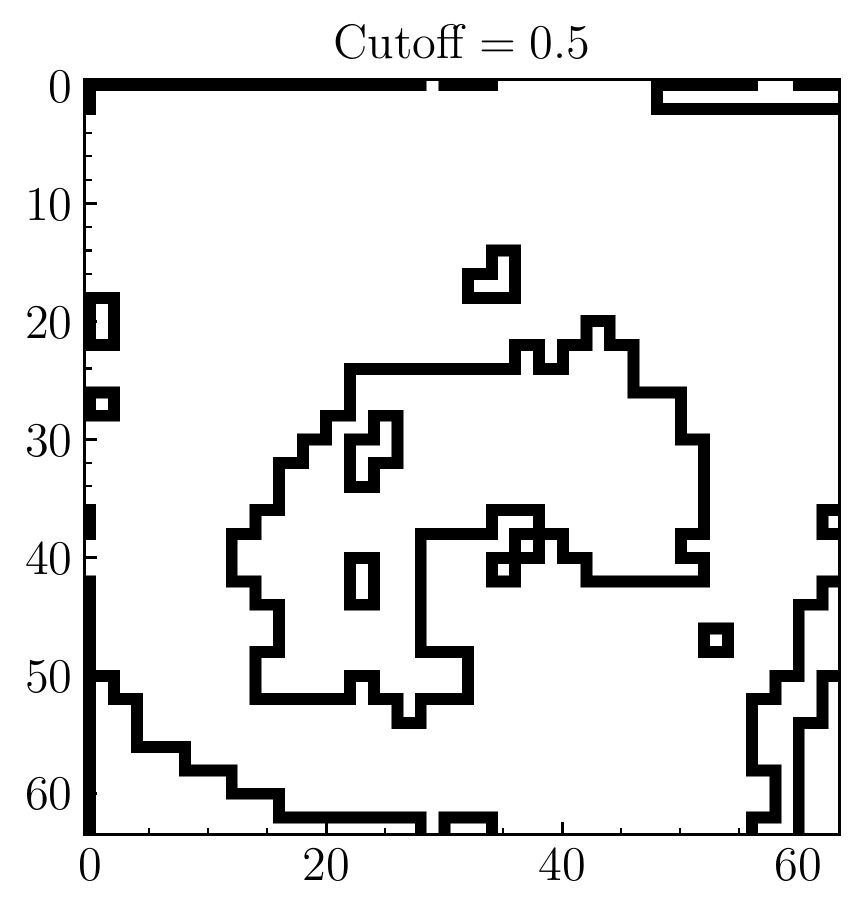}\hfill
    \includegraphics[width=0.31\textwidth]{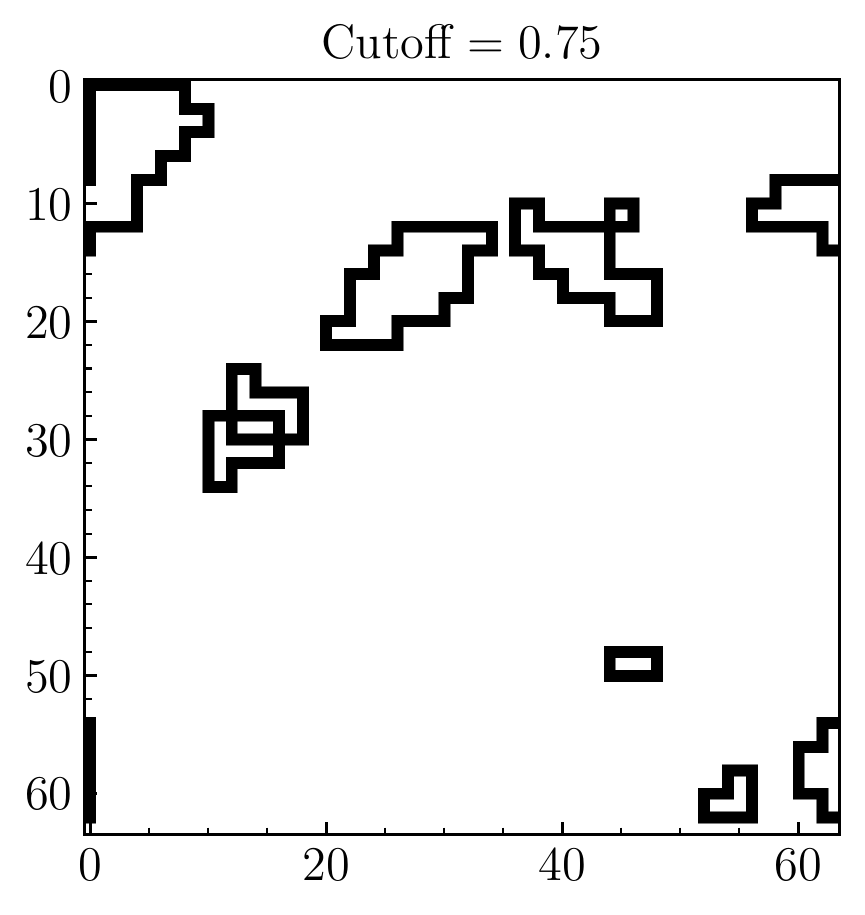}
    \caption{\label{fig:CIFAR10_preprocessing} Example of preprocessing steps
    for converting CIFAR-10 images to `worm-like' images.}
\end{figure*}

\begin{figure*}[htb]
    \centering
    \includegraphics[width=0.48\textwidth]{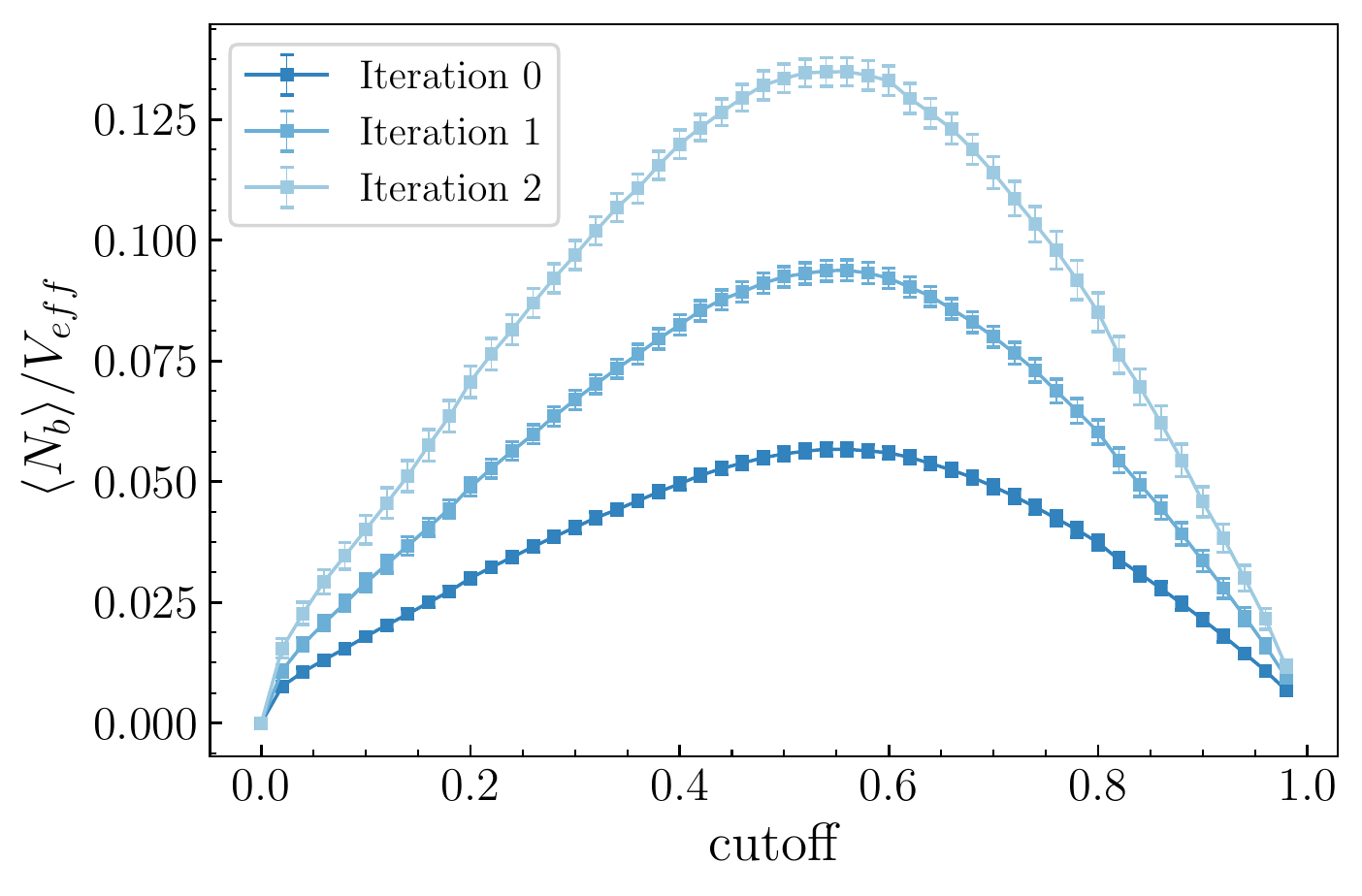}
    \includegraphics[width=0.48\textwidth]{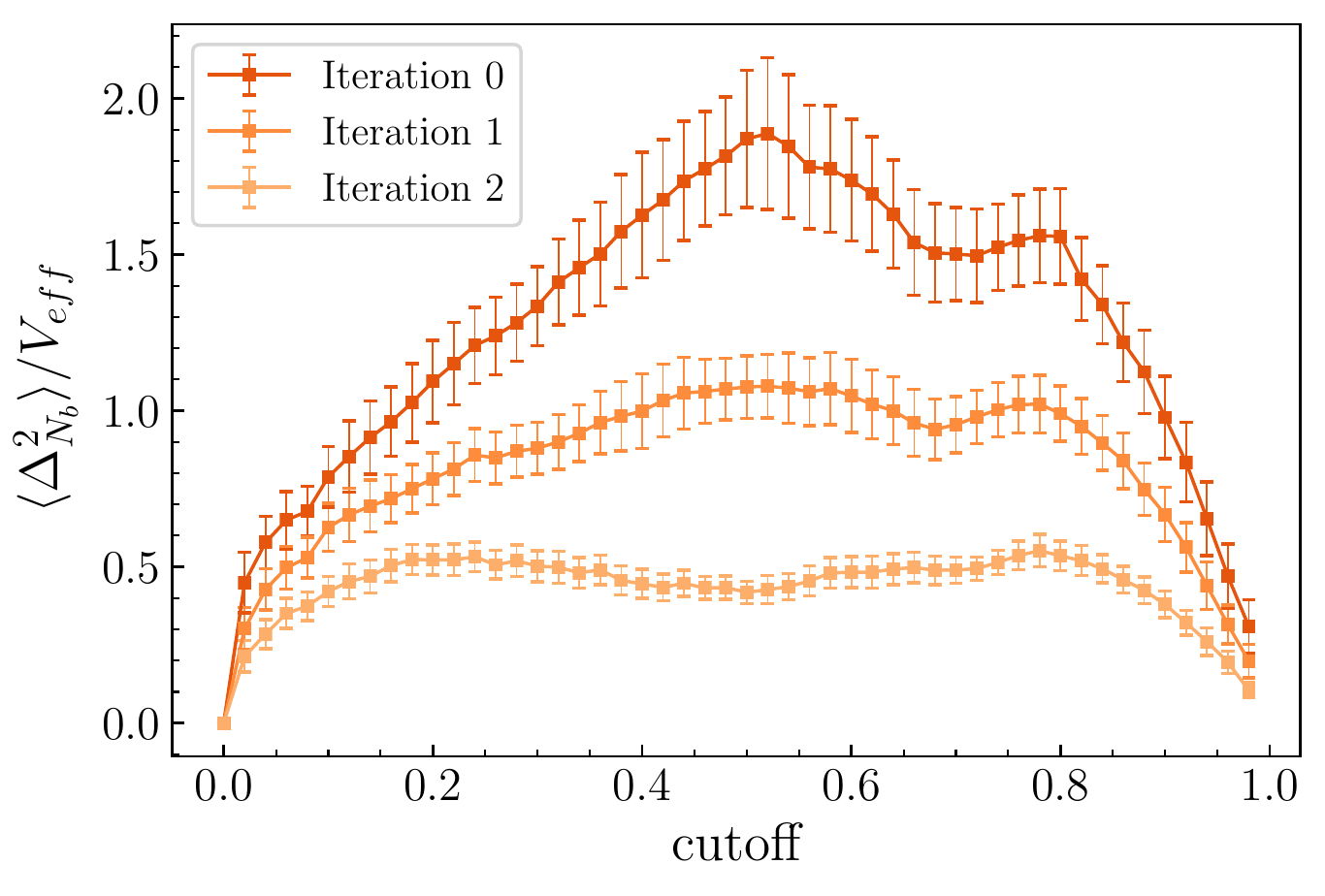}
    \caption{$\langle N_b\rangle$ and $\langle \Delta_{N_b}^2\rangle$ vs.
        grayscale cutoff value for 500 randomly chosen images from the CIFAR-10
    dataset.}
    \label{fig:CIFAR10_bond_stats}
\end{figure*}

Having better understood how these RG transformations can be used to describe
the 2D Ising model near criticality, we began to look for possible applications
to real-world datasets. For our analysis, we used the CIFAR-10
\cite{Krizhevsky09} image set consisting of $60,000$ $32\times32$ color images
in 10 classes. First, each of the images were converted to a grayscale with
pixel values in the range $[0, 1]$. Next, a grayscale cutoff value was chosen
so that all pixels with values below the cutoff would become black, and pixels
above the cutoff would become white, resulting in images consisting entirely of
black and white pixels. Finally, each of these images were converted to
`worm-like' images by drawing the boundaries separating black and white
collections of pixels. An example of these preprocessing steps are illustrated
in Fig.~\ref{fig:CIFAR10_preprocessing}.
This process was carried out on a mini-batch consisting of 500 randomly
selected images from the CIFAR-10 image set. For each image in our
mini-batch, we calculated $\langle N_b\rangle$ and $\langle
\Delta_{N_b}^2\rangle$ over a range of grayscale cutoff values in $[0, 1]$
in steps of $0.02$. Each of these images were then iteratively blocked using
the $(1 + 1 \rightarrow 0)$ blocking procedure described in Sec.~\ref{sec:trg},
calculating $\langle N_b\rangle$ and $\langle \Delta_{N_b}^2\rangle$ for each
successive blocking step, as shown in Fig.~\ref{fig:CIFAR10_bond_stats}.
Immediately we see that there is no identifiable low temperature phase, and
that for cutoff values near both $0$ and $1$, we obtain images which are mostly
empty, similar to the high temperature configurations obtained from the worm
algorithm.  This suggests that there is no such notion of criticality (as
characterized by the abrupt transition from a low to high temperature phase)
like we found for the two-dimensional Ising model.
% the extremes we obtain mostly empty images, similar
% to the high temperature phase we saw previously.
% Further, we see that $\langle N_b\rangle / V_{eff}$ attains a maximum for
% cutoff values near $0.5$, indicating that

%\par
%\listoftodos
\end{document}